\title{Who Wins the Conflict?\\Mechanistic Interpretability of Text Bias in Audio LLMs}

\author{Hyebin Cho\quad Suho Yoo\quad Jaehyuk Jang\quad Changick Kim\quad Joon Son Chung  \\[0.5em]
Korea Advanced Institute of Science and Technology, South Korea \\[0.5em]
\texttt{\{hyebin.cho, suho.yoo, jhyuk, changick, joonson\}@kaist.ac.kr}
}

\documentclass[11pt]{article}

\usepackage[preprint]{acl}

\usepackage{times}
\usepackage{latexsym}

\usepackage[T1]{fontenc}

\usepackage[utf8]{inputenc}

\usepackage{microtype}

\usepackage{inconsolata}

\usepackage{graphicx}
\usepackage{booktabs, array, multirow, amsmath, cleveref}

\usepackage{times}
\usepackage{epsfig}
\usepackage{float}
\usepackage{placeins}
\usepackage{enumitem}
\usepackage{tabularx}
\usepackage{xstring}
\usepackage{xspace}
\usepackage[hang,flushmargin]{footmisc}
\usepackage[utf8]{inputenc} 
\usepackage[T1]{fontenc}    
\usepackage{url}            
\usepackage{booktabs}       
\usepackage{amsfonts}       
\usepackage{nicefrac}       
\usepackage{xcolor}         
\usepackage{amsmath, amssymb, overpic, textpos}
\usepackage{graphicx}
\usepackage{multirow}
\usepackage{color, colortbl}
\usepackage{tabulary}
\usepackage{listings}
\usepackage{multicol}
\usepackage{xspace}
\usepackage{graphbox}
\usepackage{arydshln}
\usepackage{adjustbox}
\usepackage{kotex}
\usepackage{caption}
\usepackage{subcaption}
\usepackage{makecell}
\usepackage{tabularray}
\usepackage{bm}
\usepackage{tcolorbox}
\usepackage{etoc}
\usepackage{titletoc}
\usepackage{cleveref}

\usepackage[table]{xcolor}
\definecolor{aliceblue}{rgb}{0.94, 0.97, 1.0}

\begin{document}
\maketitle
\begin{abstract}

While Audio Large Language Models (Audio LLMs) excel at multimodal understanding, they suffer from \textit{text dominance}, a bias where models favor text over acoustic evidence, potentially leading to hallucinated responses. However, the internal mechanisms underlying how these models behave when audio and textual inputs contradict each other remain unexplored. In this work, we present the first mechanistic analysis of this phenomenon by tracing the propagation of internal representations across layers.
Our investigation reveals three key findings:
(i) text dominance is consistently observed across models; 
(ii) while text and audio rely on functionally distinct pathways, they ultimately converge into a shared semantic space in late layers; and 
(iii) the text pathway does not erase audio information, but rather actively suppresses intact audio representations.
Building on these insights, we leverage back-patching, a training-free intervention that routes late-layer audio activations back into earlier layers. This amplifies the audio representations, enabling them to overcome textual suppression. Our evaluation shows that back-patching consistently reduces text dominance, demonstrating a mechanistic route to mitigating text dominance under conflict. 

\end{abstract}

\section{Introduction}
Large Language Models (LLMs) have seamlessly integrated the audio domain, giving rise to highly capable Audio Large Language Models (Audio LLMs)~\cite{qwen2, gong2024listen, ultravox}. Advancing beyond simple speech translation, modern models directly process acoustic signals across a broad spectrum of applications---from captioning and QA~\cite{cho2025learning, ghosh2024gama} to high-dimensional reasoning tasks like multi-audio comparison, logical deduction, and robustness testing~\cite{chen2024beyond, diao2025soundmind, pan2025wild, hou2025evaluating, deshmukh2025audio, tang2024can}. 

\begin{figure*}[t]
    \centering
    \includegraphics[width=\textwidth]{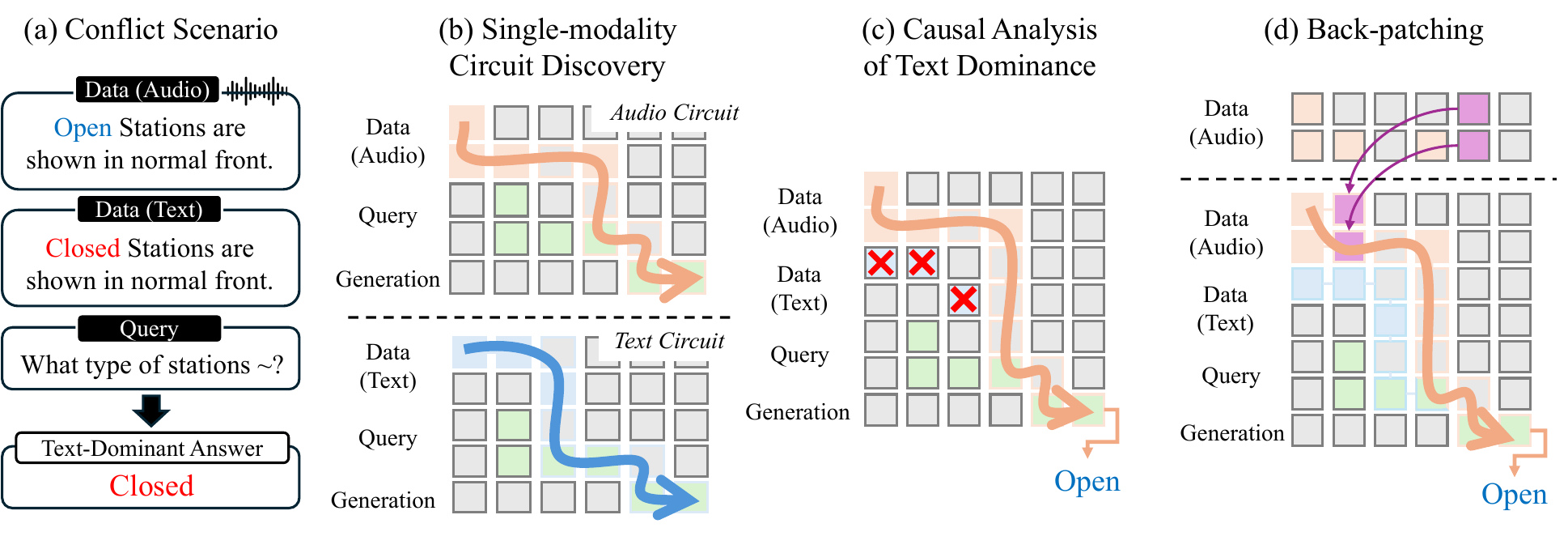}     
    \vspace{-25pt}
    \caption{\textbf{Overview of our approach to analyzing and mitigating text dominance in Audio LLMs.}
(a) An example of modality conflict where the model ignores acoustic evidence and generates a text-dominant answer.
(b) Identification of the individual text and audio circuits and the analysis of their functional relationship.
(c) Targeted ablations reveal that removing the text circuit restores the model's reliance on audio cues, implicating the text circuit as a major driver of the bias.
(d) A training-free, test-time intervention that mitigates text dominance by routing late-layer audio activations to earlier layers.}
    \vspace{-20pt}
    \label{fig:overall}
\end{figure*}

However, a key challenge undermines their reliability in practical deployment: \textit{text dominance}, a bias where models favor text over conflicting audio~\cite{weck2024muchomusic, wu2025language, alme}. For example, if a prompt contains a flawed transcript, the model may generate a response consistent with the textual premise despite clear contradictory evidence in the speech. \cref{fig:overall}(a) illustrates a typical instance of this failure mode, where the model overlooks clear acoustic evidence and incorrectly generates a text-aligned response.

Mitigating this vulnerability requires uncovering how competing text and audio representations interact beneath the surface. Consequently, Mechanistic Interpretability (MI) emerges as a natural tool for tracing these internal multimodal interactions. Yet, while recent MI studies have begun to explore information flow in Audio LLMs~\cite{chowdhury2026ar, pluth2026mechanistic, chen2026causal, glazer2026audio}, they have primarily focused on cooperative multi-modal setups or single-modality tracking. The mechanisms governing how text and audio representations interact under conflicting conditions remain entirely unexplored.

In this work, we present the first mechanistic interpretation of text-audio conflict in Audio LLMs at the circuit level, tracing how modality-specific signals propagate through attention heads and MLP modules. To achieve this, we first reconstruct a dataset across four diverse tasks, where each sample features conflicting text and audio prompts (\cref{fig:overall}a). We then independently isolate text and audio circuits (\cref{fig:overall}b), and perform causal ablation to capture how these modalities interact under conflict (\cref{fig:overall}c). Our comprehensive analysis uncovers the internal circuits underlying multimodal conflict in representative open-source Audio LLMs.



Our analysis yields three key findings on text-audio 
conflict in Audio LLMs. First, we empirically demonstrate a strong tendency towards text dominance, with models favoring text over conflicting acoustic evidence across~(\S\ref{subsec:dataset}). Second, despite structural overlaps, text and audio circuits remain functionally independent during data processing, yet their activations ultimately align in late layers~(\S\ref{subsec:circuit}). Third, text dominance does not arise from the erasure of audio information; rather, acoustic evidence remains recoverable even when the text pathway dominates~(\S\ref{subsec:causal}).

Building on these observations, we adopt back-patching~\cite{biran2024hopping, lepori2025racing, same} for the audio domain~(\cref{fig:overall}d). Since multimodal integration occurs in later layers~\cite{chen2026causal}, injecting deep acoustic features back into early stages allows us to pre-activate the audio signal. This increased activation magnitude enables the audio signal to counteract text dominance and preserve its information throughout the final fusion layers.
Evaluated across eight languages and four conflict types, our back-patching intervention elevates the average audio accuracy from 0.35 to 0.46 without compromising the underlying text circuit, approaching the 0.50 balanced reference in our conflict setting

\section{Related Work}

\subsection{Audio Large Language Models and Modality Bias}


While modern Audio LLMs have shown remarkable performance across various downstream tasks, they exhibit a critical vulnerability known as modality bias~\cite{weck2024muchomusic, wu2025language, alme}. Heavily optimized for language priors, these models often demonstrate \textit{text dominance}—disproportionately favoring textual prompts over explicit but contradictory acoustic evidence. This phenomenon aligns with the strict modality hierarchy observed in broader Omni-LLMs, where audio is frequently suppressed by dominant visual or textual signals~\cite{yan2026beyond, sung2025avhbench, Zhang_2026_CVPR, wen2026vision}. However, while cross-modal conflicts are actively addressed in Omni-LLMs, research on dedicated Audio LLMs remains sparse. Prior mitigation has primarily focused on training-based approaches. In contrast, we focus on mechanistically characterizing text dominance under conflict and derive back-patching as a training-free intervention from this analysis.

\subsection{Mechanistic Interpretability}

Circuit discovery in mechanistic interpretability (MI) aims to reverse-engineer neural networks into human-understandable computational subgraphs, or circuits~\cite{geva2021transformer, nikankin2025arithmetic, conmy2023towards}. Traditional MI techniques such as causal tracing and activation patching~\cite{meng2022rome, wang2022interpretability} enable the precise identification of these circuits by directly measuring the effect of localized counterfactual interventions. To accelerate this exhaustive patching process, we adapt attribution patching with integrated gradients (AP-IG)~\cite{nanda2022attributionpatching, sundararajan2017axiomatic, hanna2024have} to the text-audio domain, allowing us to efficiently localize the distributed causal pathways where modality arbitration under conflict occurs.

Early MI efforts in Audio LLMs have begun to explore internal mechanics by employing SAEs to disentangle features~\cite{chowdhury2026ar, pluth2026mechanistic}, mapping fusion dynamics via causal tracing~\cite{chen2026causal}, or steering audio reliance~\cite{glazer2026audio}. However, despite these advancements, existing research has yet to causally explain how models resolve direct modality conflicts. We bridge this critical gap by demonstrating, through causal circuit analysis, that text dominance is actively driven by suppression from the text circuit. Furthermore, we leverage these insights to propose back-patching, a training-free intervention that mitigates this suppression by routing late-layer audio activations back into earlier layers.

\begin{table}[!t]
\centering
\small
\caption{\textbf{Examples of conflict tasks.} The table illustrates the four types of linguistic swaps (Adjective, Negation, Number, and Time) applied to base prompts, along with their corresponding counterfactual (CF) queries and choices. The correct answer is always included within the options, with the order of choices (A and B) randomized per sample.}
\vspace{-10pt}
\renewcommand{\arraystretch}{1.2}
\begin{tabular}{@{} p{0.95\columnwidth} @{}}
\toprule
\textbf{1. Adjective Swap} \\
\quad \textit{Base:} \textbf{Closed} stations are shown in normal font. \\
\quad \textit{CF:} \textbf{Open} stations are shown in normal font. \\
\quad \textit{Query:} What type of stations are shown in normal font? \\ 
\quad \textit{Choices:} A: Open $\mid$ B: Closed \\
\midrule
\textbf{2. Negation Swap} \\
\quad \textit{Base:} My attorney \textbf{didn't recommend} that I take\dots \\
\quad \textit{CF:} My attorney \textbf{recommended} that I take\dots \\
\quad \textit{Query:} Did the attorney advise taking...? \\
\quad \textit{Choices:} A: Recommended $\mid$ B: Didn't Recommend\\
\midrule
\textbf{3. Number Swap} \\
\quad \textit{Base:} Stoppard has been married \textbf{three times}. \\

\quad \textit{CF:} Stoppard has been married \textbf{two times}. \\
\quad \textit{Query:} How many times has Stoppard been married? \\ 
\quad \textit{Choices:} A: Three $\mid$ B: Two \\
\midrule
\textbf{4. Time Swap} \\
\quad \textit{Base:} We've been laughing \textbf{all day} about that. \\
\quad \textit{CF:} We've been laughing \textbf{all week} about that. \\
\quad \textit{Query:} What length of time is mentioned? \\ 
\quad \textit{Choices:} A: Day $\mid$ B: Week \\
\bottomrule
\end{tabular}
\label{tab:counterfactual_prompts_1col}
\end{table}

\begin{table}[!t]
\centering
\small
\caption{\textbf{Modality preference under base and counterfactual (CF) conditions.} $X\Rightarrow Y$ denotes the prompt sequence order where $X$ precedes $Y$. Due to the binary choice format, $\Delta_{T-A}$ explicitly captures the accuracy gap between the text and audio modalities ($\text{Text Acc.} - \text{Audio Acc.}$). Bold values highlight a prominent text bias ($\Delta_{T-A} > 0.1$).}
\vspace{-10pt}
\begin{tabular}{@{} l cc cc @{}}
\toprule
\multirow{2}{*}{\textbf{Condition}} & \multicolumn{2}{c}{\textbf{Qwen}} & \multicolumn{2}{c}{\textbf{Ultravox}} \\
\cmidrule(lr){2-3} \cmidrule(lr){4-5}
& Text Acc. & $\Delta_{T-A}$ & Text Acc. & $\Delta_{T-A}$ \\
\midrule
$A_{base} \Rightarrow T_{CF}$ & 0.71 & \textbf{+0.41} & 0.56 & \textbf{+0.12} \\
$A_{CF} \Rightarrow T_{base}$ & 0.70 & \textbf{+0.39} & 0.62 & \textbf{+0.25} \\
$T_{base} \Rightarrow A_{CF}$ & 0.50 &  0.00 & 0.66 & \textbf{+0.32} \\
$T_{CF} \Rightarrow A_{base}$ & 0.47 & $-0.05$ & 0.61 & \textbf{+0.22} \\
\bottomrule
\end{tabular}
\vspace{-15pt}
\label{tab:motivation_subscript}
\end{table}

\section{The Text Dominance Phenomenon in Modality Conflict}
\label{subsec:dataset}
In this section, we first reconstruct the evaluation setup into a controlled modality conflict benchmark designed to isolate text-audio bias in Audio LLMs. We then demonstrate that representative Audio LLMs consistently exhibit strong text dominance under conflicting multimodal inputs.

\subsection{Task Formulation and Dataset Construction}



To evaluate modality bias, we reformulate a modality conflict task based on the ALME benchmark~\cite{alme}, where audio and text inputs contain explicitly contradictory facts. As shown in~\cref{tab:counterfactual_prompts_1col}, the original ALME prompts consist of base and counterfactual (CF) pairs created by flipping specific attributes (e.g., text: "Closed", audio: "Open").
To ensure that this evaluation reflects genuine modality preference rather than dataset or sequence-related artifacts, we further regenerate the corresponding audio prompts using a text-to-speech module~\footnote{\url{https://github.com/rany2/edge-tts}}. We additionally evaluate two input orders: audio-first ($A \Rightarrow T$) and text-first ($T \Rightarrow A$). Together, these design choices allow us to construct balanced text-audio conflict pairs across both modalities while controlling for modality-ordering effects. Detailed dataset construction procedures are provided in~\cref{appx:infer_dataset_construction}.

In this conflict setting, onsistently favoring one modality indicates a  modality bias. We therefore use \textit{Modality Equilibrium} as a diagnostic reference, corresponding to equal rates of following conflicting textual and acoustic evidence. Under our balanced binary conflict construction, this corresponds to 0.5 text accuracy and 0.5 audio accuracy.

\begin{figure*}[!t]
    \centering
    \includegraphics[width=\textwidth]{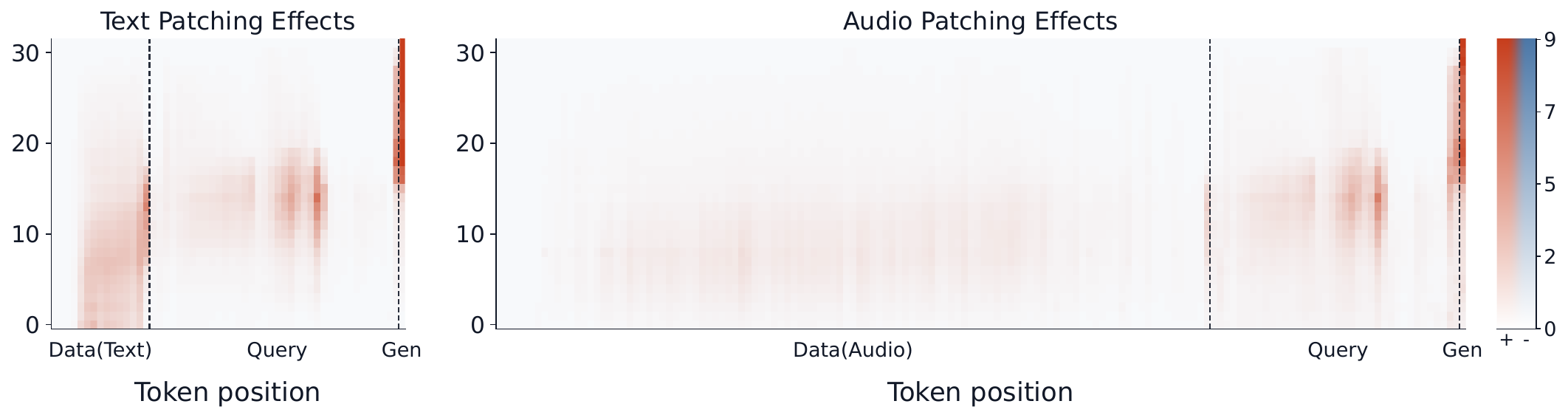}
    \vspace{-20pt}
    \caption{\textbf{Causal tracing of model components via AP-IG.} We visualize the patching effects from all model components across each input position and layer using Qwen on the negation swap conflict task. Distinct operational behaviors are observed across different functional segments, corresponding to the specific positions of Data, Query, and Generation tokens.}
    \label{fig:patching}
    \vspace{-15pt}
\end{figure*}

\subsection{Observation of Modality Bias} 
We evaluate modality conflict using two representative Audio LLMs: Qwen2-Audio-7B-Instruct and Ultravox v0.6-llama-3.1-8b. As shown in~\cref{tab:motivation_subscript}, both models deviate substantially from this balanced reference, exhibiting strong text dominance.

Specifically, Qwen exhibits an asymmetric modality pull. When text is presented last ($A \Rightarrow T$), it exploits the recency advantage and strongly dominates the output, reaching nearly 0.70 text accuracy. In contrast, when audio is presented last ($T \Rightarrow A$), the effect is weaker and less consistent; in one condition ($T_{CF} \Rightarrow A_{base}$), text accuracy even drops to 0.47, indicating a slight shift toward audio. By contrast, Ultravox maintains a stable text bias above 0.60 across all conditions, revealing a modality preference that is largely invariant to input order.

While these metrics confirm behavioral text bias, they treat the model as a black box. To uncover where and how textual features actively suppress audio during the fusion process, we must shift from surface-level observations to mechanistic interpretability, isolating the specific computational circuits driving this phenomenon.


\section{Disentangling Text and Audio Circuits} 
\label{subsec:circuit}

To map the internal landscape of Audio LLMs during modality conflict, we must first isolate the independent computational subgraphs (attention heads and MLPs) responsible for processing each modality. Following the disentanglement framework of \citet{same}, we adapt the TransformerLens library \cite{nanda2022transformerlens} for continuous audio inputs, enabling us to extract these modality-specific circuits and analyze their architectural distribution.

\subsection{Circuit Discovery via Activation Patching}
We start by identifying the model components critical for each single modality. All patching experiments are conducted in English to maintain structurally identical counterfactual pairs. To assess causal effects, we employ activation patching. For a given target modality, we define a clean run using the base prompt (containing the original factual statement) and a corrupted run using the corresponding counterfactual (CF) prompt (where the key factual information is inverted). To scale this causal estimation efficiently across all components, we utilize AP-IG \cite{sundararajan2017axiomatic, hanna2024have, same}. Instead of performing prohibitively expensive exhaustive interventions, AP-IG estimates hookwise causal importance via a path-integrated gradient attribution score:
\begin{equation}
\mathcal{S}_{IG}(v) = \left( a_v^{(clean)} - a_v^{(cf)} \right)^T \cdot \frac{1}{K} \sum_{k=1}^K \nabla_{a_v} \mathcal{F}(\tilde{e}_k)
\end{equation}
where $a_v^{(clean)}$ and $a_v^{(cf)}$ represent the activations of component $v$ under the clean and counterfactual conditions, respectively. The gradient of our target metric $\mathcal{F}$ (the logit difference) is evaluated at interpolated input embeddings $\tilde{e}_k = e^{(cf)} + \frac{k}{K}(e^{(clean)} - e^{(cf)})$. Following \cite{hanna2024have, same}, we set $K=5$. We compute these scores for each discovery prompt and average them. We then construct candidate circuits by selecting the top-ranked attention heads and MLP neurons at multiple sparsity levels $p \in \{0.1\%, \dots, 100\%\}$ based on the absolute AP-IG scores. Additional implementation details, including the discovery/evaluation split and prompt setup, are provided in~\cref{appdx:circuit_setup}.

Based on these computed AP-IG scores, \cref{fig:patching} visualizes the causal patching effects for Qwen. The heatmaps reveal distinct modality patterns: in the Data (D) positions, textual components exhibit notably stronger causal importance that extends deeper into the model layers compared to the audio components. Conversely, the activation patterns in the Query (Q) and Generation (G) positions appear remarkably similar between the two modalities.

\subsection{Circuit Faithfulness Evaluation}
\begin{figure}[t]
    \centering
    \includegraphics[width=\linewidth]{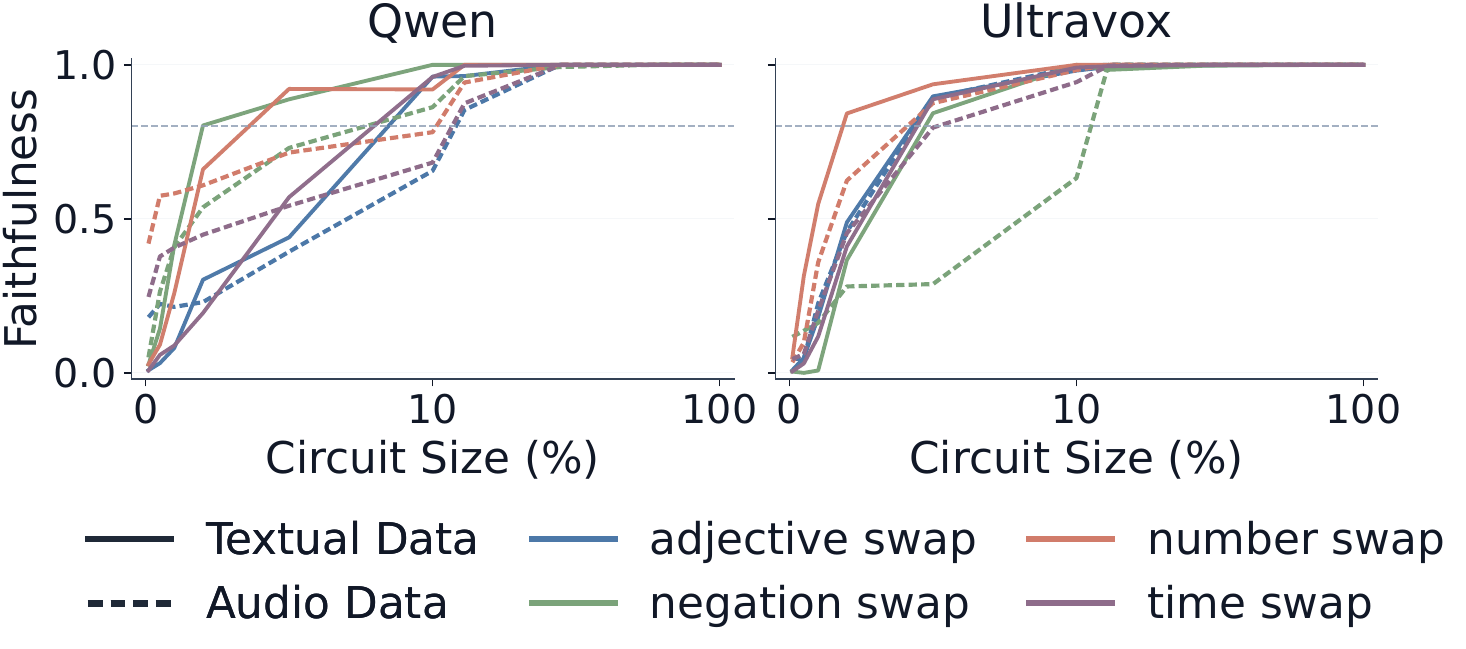}
    \vspace{-20pt}
    \caption{\textbf{Faithfulness of discovered circuits.} We evaluate circuit faithfulness across models and English conflict tasks. The smallest circuit with faithfulness above 0.8 is used for subsequent analysis.}
    \vspace{-15pt}
    \label{fig:faithfulness}
\end{figure}

To evaluate faithfulness, we retain only the selected circuit components and ablate all non-circuit components by replacing their activations with those from the corresponding counterfactual run. We sweep over circuit sparsity levels and select the smallest percentile whose normalized faithfulness exceeds 0.8. As illustrated in the Faithfulness Graphs (\cref{fig:faithfulness}), relatively sparse circuits recover a substantial portion of the original performance; this criterion yields 20\% for Qwen and 10\% for Ultravox for subsequent analysis. Notably, at equivalent sparsity levels (e.g., $p=0.1$), the audio circuits generally exhibit lower faithfulness than the text circuits, suggesting that audio processing is more distributed and therefore requires a larger fraction of components to recover comparable performance. 

\subsection{Structural and Functional Overlap Analysis} 
\label{circuit_overlap}
\begin{figure}[!t]
    \centering
    \includegraphics[width=\linewidth]{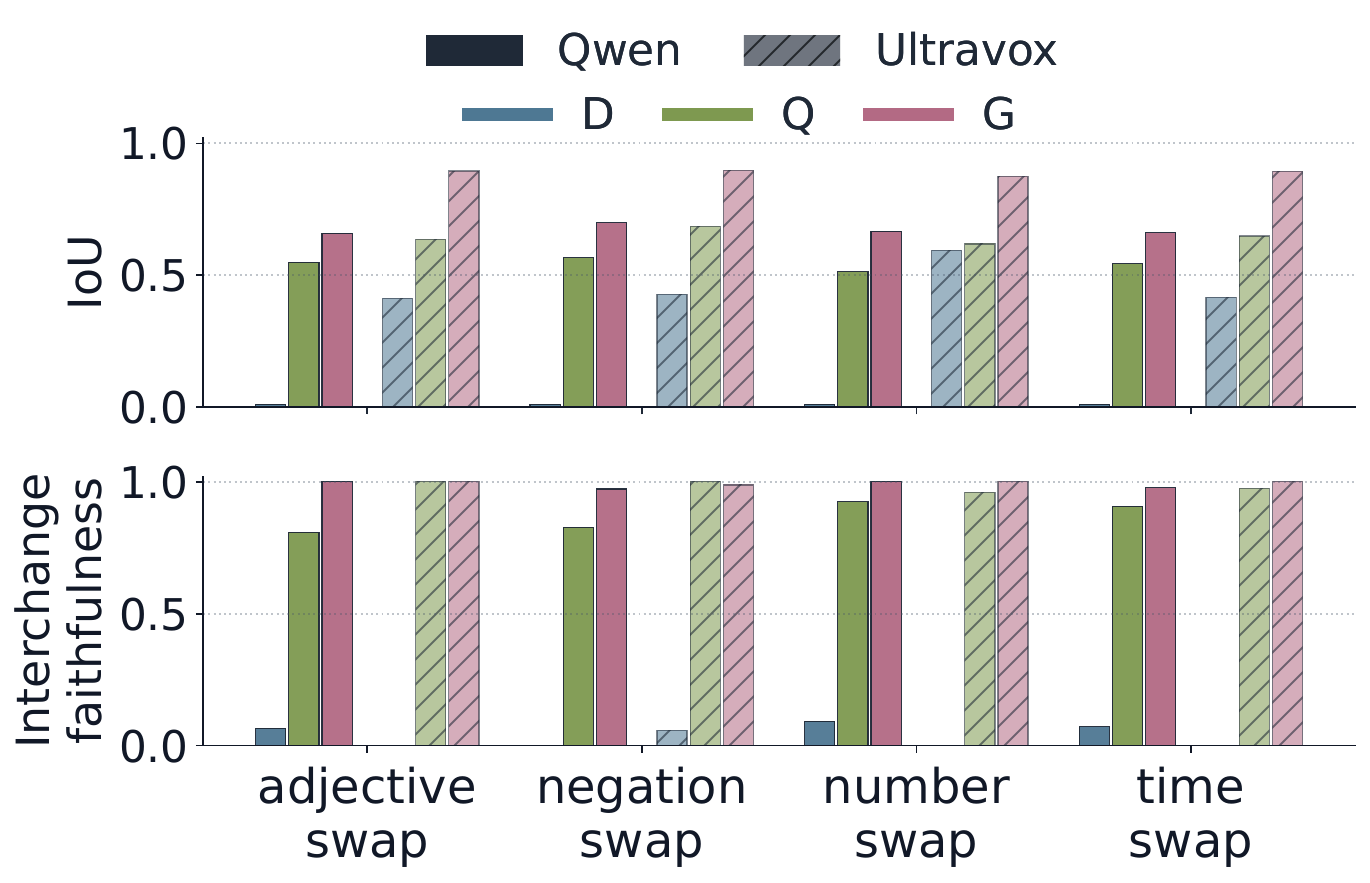}
    \vspace{-20pt}
    \caption{\textbf{Structural intersection and functional interchange of text and audio circuits.} The bar charts display the normalized Intersection over Union (IoU) and interchange faithfulness across three functional positions: data (D), query (Q), and generation (G). Qwen is represented by solid bars, while Ultravox is distinguished by hatched patterns.}
    \label{fig:iou}
    \vspace{-15pt}
\end{figure}

Given the faithful textual and audio circuits, we investigate their intersection.~\cite{kaduri2025s, same} Guided by our patching effect observations, we partition the circuits into three functional positions: Data ($c_D$), Query ($c_Q$), and Generation ($c_G$). We then evaluate the modality sub-circuits at each position for both structural overlap and functional equivalence (\cref{fig:iou}).

\begin{table*}[!t]
\centering
\small
\caption{\textbf{Audio accuracy under different ablation settings.} In this closed-set A/B evaluation, predictions are restricted to the two candidate labels, so text accuracy is exactly $1 - \text{audio accuracy}$. Removing the text circuit ($-\mathcal{C}_T$) shifts model reliance toward audio, whereas removing the audio circuit ($-\mathcal{C}_A$) further reduces audio accuracy by reducing audio-consistent predictions under conflict. Ablating the union circuit ($-\mathcal{C}_U$) shifts the models toward the 0.5 balanced reference; however, Ultravox remains more text-dominant than Qwen, indicating stronger influence from its background circuits.}
\setlength{\tabcolsep}{4pt} 
\resizebox{\textwidth}{!}{
\begin{tabular}{l cccc cccc cccc cccc}
\toprule
\multirow{2}{*}{\textbf{Model}} & \multicolumn{4}{c}{\textbf{Adjective}} & \multicolumn{4}{c}{\textbf{Negation}} & \multicolumn{4}{c}{\textbf{Number}} & \multicolumn{4}{c}{\textbf{Time}} \\
\cmidrule(lr){2-5} \cmidrule(lr){6-9} \cmidrule(lr){10-13} \cmidrule(lr){14-17}
& Full & $-\mathcal{C}_T$ & $-\mathcal{C}_A$ & $-\mathcal{C}_U$ & Full & $-\mathcal{C}_T$ & $-\mathcal{C}_A$ & $-\mathcal{C}_U$ & Full & $-\mathcal{C}_T$ & $-\mathcal{C}_A$ & $-\mathcal{C}_U$ & Full & $-\mathcal{C}_T$ & $-\mathcal{C}_A$ & $-\mathcal{C}_U$ \\
\midrule
\textbf{Qwen}  & 0.33 & 0.82 & 0.13 & 0.45 & 0.40 & 0.67 & 0.45 & 0.55 & 0.44 & 0.85 & 0.13 & 0.45 & 0.37 & 0.79 & 0.16 & 0.50 \\
\midrule
\textbf{Ultra.} & 0.37 & 0.92 & 0.04 & 0.39 & 0.22 & 0.81 & 0.06 & 0.44 & 0.18 & 0.95 & 0.00 & 0.35 & 0.36 & 0.93 & 0.01 & 0.40 \\
\bottomrule
\end{tabular}
}
\label{tab:ablation_transition}
\end{table*}

\noindent\textbf{Structural intersection.} To quantify the overlap between the textual ($\mathcal{C}_T$) and audio ($\mathcal{C}_A$) circuits, we directly calculate the Intersection over Union (IoU) within their shared positional space. To correct for the bias of larger circuits inherently yielding higher overlaps, we normalize the metric against randomly sampled circuits ($c_T^{rand}$, $c_A^{rand}$) of identical sizes. We define the normalized IoU (NIoU) as:
\begin{equation}
\text{NIoU}(\mathcal{C}_T, \mathcal{C}_A) = \frac{\text{IoU}(\mathcal{C}_T, \mathcal{C}_A) - \text{IoU}(\mathcal{C}_T^{rand}, \mathcal{C}_A^{rand})}{1.0 - \text{IoU}(\mathcal{C}_T^{rand}, \mathcal{C}_A^{rand})}
\end{equation}
In Qwen, data sub-circuits are nearly disjoint (IoU $\approx$ 0.01), indicating independent early processing channels. In contrast, Ultravox exhibits early structural entanglement (data IoU $> 0.40$). Despite these initial differences, both models show substantially greater overlap at the query (IoU 0.54--0.68) and generation (IoU 0.66--0.89) stages.




\noindent\textbf{Functional interchange.} Structurally disjoint components may still implement similar functionality. We therefore investigate whether corresponding sub-circuits are functionally interchangeable. For example, we test whether activating the audio query sub-circuit ($\mathcal{C}_A^Q$) during a purely textual forward pass---effectively replacing its textual counterpart ($\mathcal{C}_T^Q$)---preserves faithfulness. To measure this equivalence, we define the average Interchange Faithfulness (IF):
\begin{equation}
\begin{aligned}\text{IF}(Q) = \frac{1}{2} \Big(& \mathcal{S}(\mathcal{C}_T^D \cup \mathcal{C}_{A \rightarrow T}^Q \cup \mathcal{C}_T^G, \mathcal{T}_T) \\
& + \mathcal{S}(\mathcal{C}_A^D \cup \mathcal{C}_{T \rightarrow A}^Q \cup \mathcal{C}_A^G, \mathcal{T}_A)
\Big)
\end{aligned}
\end{equation}
where $\mathcal{S}(\cdot, \mathcal{T})$ denotes the faithfulness score evaluated on the respective uni-modal task prompt ($\mathcal{T}_T$ for text, $\mathcal{T}_A$ for audio). We show IF($Q$) as a representative case; IF($D$) and IF($G$) are defined analogously. For the Data position, the interchange is defined using the modality-specific data-token span rather than exact token-index identity. Swapping data sub-circuits degrades performance to near-random baselines, whereas interchanging query and generation sub-circuits preserves performance close to the original uni-modal circuits. This suggests that by the query and generation stages, the two modalities operate in a shared semantic space, which we verify in the following section.

\begin{figure}[t] 
    \centering
    \includegraphics[width=\linewidth]{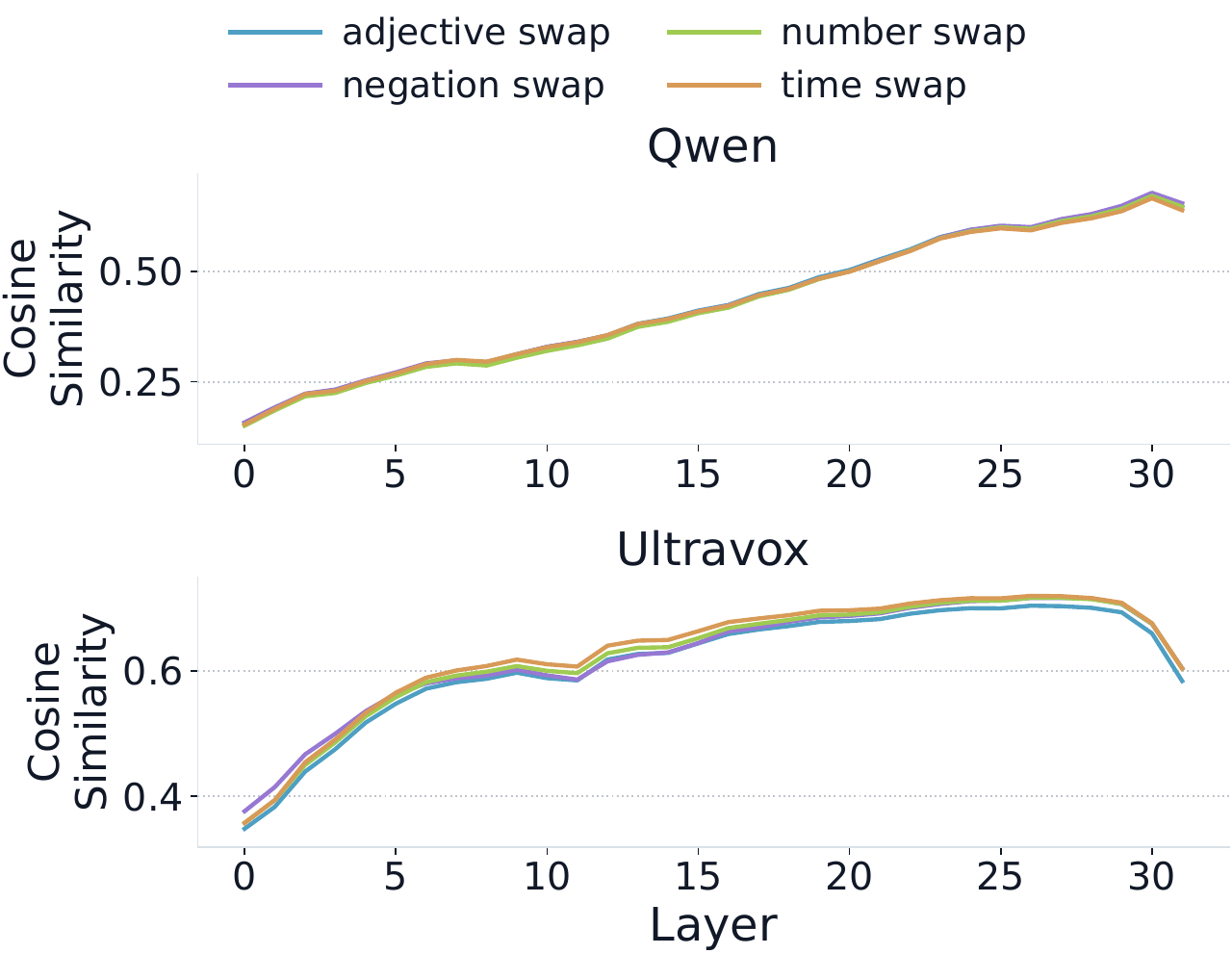}
    \vspace{-20pt}
    \caption{\textbf{Cosine similarity of audio and textual token activations across layers.} For each audio token, we compute its maximum cosine similarity against all corresponding textual tokens, then average the scores. The similarity consistently increases in deeper layers of both models, indicating that acoustic features align with textual representations late in the network.}
    \label{fig:token_activation_combined}
    \vspace{-15pt}
\end{figure}

\subsection{Semantic Convergence in Deep Layers}

To understand how these initially distinct modalities eventually converge, we further investigate how their representations evolve across the network. Unlike the preceding circuit-level analysis over data, query, and generation positions, this analysis compares token-level hidden representations within the modality-specific data spans. Specifically, within the modality-specific data spans, we compute cosine similarity at each layer without assuming one-to-one token alignment, taking for each token in one modality the maximum cosine similarity over all tokens in the other modality and then averaging the resulting scores. As illustrated in \cref{fig:token_activation_combined}, the similarity scores consistently increase in deeper layers for both Qwen and Ultravox. This observation suggests that audio representations ultimately converge into a shared semantic space with text in the late layers.


\section{Causal Ablation of Text Dominance}
\label{subsec:causal}
In this section, we analyze how textual and audio circuits interact during explicit modality conflict. Following our causal ablation design (\cref{tab:ablation_transition}), we reveal that text dominance is actively driven by the text circuit suppressing acoustic signals. Finally, we show that brute-force circuit deletion degrades core language capabilities, demonstrating the necessity for a non-destructive intervention.



\subsection{Causal Ablation Design} First, we investigate the root cause of text dominance by selectively deactivating the text ($\mathcal{C}_T$), audio ($\mathcal{C}_A$), and union ($\mathcal{C}_U$) circuits. To precisely isolate their causal effects, we perform position-specific mean ablation exclusively at the respective token positions: text tokens ($\mathcal{P}_T$) for the text circuit, audio tokens ($\mathcal{P}_A$) for the audio circuit, and both for the union circuit. We evaluate the models in a closed-set classification setting where they must choose between audio-supported and text-supported labels. For text-circuit ablation, we replace the activations of all components in $\mathcal{C}_T$ at text-token positions with their baseline mean activations: \begin{equation}a_c[p] \leftarrow \bar{a}_c \quad \forall c \in \mathcal{C}_T,\; p \in \mathcal{P}_T\end{equation} where $\bar{a}_c$ is the mean activation of component $c$ estimated over a balanced baseline distribution to avoid out-of-distribution (OOD) effects. We apply the same procedure to $\mathcal{C}_A$ and $\mathcal{C}_U$ at $\mathcal{P}_A$ and $\mathcal{P}_T \cup \mathcal{P}_A$, respectively.

\begin{table*}[t]
\centering
\small
\caption{\textbf{Back-patching performance across conflicting modality conditions.} Results are averaged over two complementary setups: (i) counterfactual audio with literal text, and (ii) literal audio with counterfactual text. We report the patched audio accuracy and the bootstrap improvement ($\Delta$, over the unpatched baseline) with standard deviations in parentheses. An audio accuracy of 0.5 denotes the balanced reference point under our conflict setting.}
\scalebox{0.94}{
\begin{tabular}{ll c@{\hspace{2pt}}c@{\hspace{8pt}} c@{\hspace{2pt}}c@{\hspace{8pt}} c@{\hspace{2pt}}c@{\hspace{8pt}} c@{\hspace{2pt}}c@{\hspace{8pt}} c@{\hspace{2pt}}c}
\toprule
\multirow{2}{*}{\textbf{Model}}
& \multirow{2}{*}{\textbf{Lang.}}
& \multicolumn{2}{c}{\textbf{Adj. Swap}}
& \multicolumn{2}{c}{\textbf{Neg. Swap}}
& \multicolumn{2}{c}{\textbf{Num. Swap}}
& \multicolumn{2}{c}{\textbf{Time Swap}}
& \multicolumn{2}{c}{\textbf{Avg.}}\\
\cmidrule(lr){3-4} \cmidrule(lr){5-6} \cmidrule(lr){7-8} \cmidrule(lr){9-10} \cmidrule(lr){11-12}
& & Patched & $\Delta$ & Patched & $\Delta$ & Patched & $\Delta$ & Patched & $\Delta$ & Patched & $\Delta$\\
\midrule
\multirow{9}{*}{Qwen}
& AR   & 0.46 & $+0.30_{\scriptscriptstyle \pm 0.09}$ & 0.53 & $+0.20_{\scriptscriptstyle \pm 0.09}$ & 0.47 & $+0.17_{\scriptscriptstyle \pm 0.10}$ & 0.45 & $+0.23_{\scriptscriptstyle \pm 0.10}$ & \cellcolor{aliceblue}0.48 & \cellcolor{aliceblue}$+0.23_{\scriptscriptstyle \pm 0.10}$ \\
& DE   & 0.51 & $+0.31_{\scriptscriptstyle \pm 0.09}$ & 0.51 & $+0.22_{\scriptscriptstyle \pm 0.11}$ & 0.48 & $+0.11_{\scriptscriptstyle \pm 0.10}$ & 0.46 & $+0.21_{\scriptscriptstyle \pm 0.10}$ & \cellcolor{aliceblue}0.49 & \cellcolor{aliceblue}$+0.21_{\scriptscriptstyle \pm 0.10}$ \\
& EN   & 0.33 & $+0.08_{\scriptscriptstyle \pm 0.04}$ & 0.42 & $+0.14_{\scriptscriptstyle \pm 0.05}$ & 0.45 & $+0.09_{\scriptscriptstyle \pm 0.04}$ & 0.39 & $+0.08_{\scriptscriptstyle \pm 0.05}$ & \cellcolor{aliceblue}0.40 & \cellcolor{aliceblue}$+0.10_{\scriptscriptstyle \pm 0.05}$ \\
& FR   & 0.46 & $+0.22_{\scriptscriptstyle \pm 0.08}$ & 0.49 & $+0.20_{\scriptscriptstyle \pm 0.08}$ & 0.50 & $+0.07_{\scriptscriptstyle \pm 0.08}$ & 0.48 & $+0.22_{\scriptscriptstyle \pm 0.08}$ & \cellcolor{aliceblue}0.48 & \cellcolor{aliceblue}$+0.18_{\scriptscriptstyle \pm 0.08}$ \\
& IT   & 0.47 & $+0.29_{\scriptscriptstyle \pm 0.08}$ & 0.50 & $+0.24_{\scriptscriptstyle \pm 0.08}$ & 0.54 & $+0.23_{\scriptscriptstyle \pm 0.09}$ & 0.49 & $+0.13_{\scriptscriptstyle \pm 0.09}$ & \cellcolor{aliceblue}0.50 & \cellcolor{aliceblue}$+0.22_{\scriptscriptstyle \pm 0.09}$ \\
& JA   & 0.38 & $+0.08_{\scriptscriptstyle \pm 0.04}$ & 0.47 & $+0.12_{\scriptscriptstyle \pm 0.10}$ & 0.47 & $+0.04_{\scriptscriptstyle \pm 0.03}$ & 0.48 & $+0.16_{\scriptscriptstyle \pm 0.08}$ & \cellcolor{aliceblue}0.45 & \cellcolor{aliceblue}$+0.10_{\scriptscriptstyle \pm 0.06}$ \\
& PT   & 0.48 & $+0.15_{\scriptscriptstyle \pm 0.09}$ & 0.48 & $+0.18_{\scriptscriptstyle \pm 0.08}$ & 0.52 & $+0.05_{\scriptscriptstyle \pm 0.09}$ & 0.48 & $+0.20_{\scriptscriptstyle \pm 0.08}$ & \cellcolor{aliceblue}0.49 & \cellcolor{aliceblue}$+0.15_{\scriptscriptstyle \pm 0.09}$ \\
& ZH   & 0.32 & $+0.08_{\scriptscriptstyle \pm 0.05}$ & 0.37 & $+0.11_{\scriptscriptstyle \pm 0.05}$ & 0.45 & $+0.06_{\scriptscriptstyle \pm 0.03}$ & 0.38 & $+0.07_{\scriptscriptstyle \pm 0.04}$ & \cellcolor{aliceblue}0.38 & \cellcolor{aliceblue}$+0.08_{\scriptscriptstyle \pm 0.04}$ \\
\cmidrule(lr){2-12}
\rowcolor{aliceblue}
& \textbf{Avg.} & 0.43 & $+0.19_{\scriptscriptstyle \pm 0.07}$ & 0.46 & $+0.16_{\scriptscriptstyle \pm 0.08}$ & 0.49 & $+0.10_{\scriptscriptstyle \pm 0.07}$ & 0.45 & $+0.16_{\scriptscriptstyle \pm 0.08}$ & 0.46 & $+0.16_{\scriptscriptstyle \pm 0.08}$ \\ 
\midrule
\multirow{9}{*}{Ultra.}
& AR   & 0.48 & $+0.10_{\scriptscriptstyle \pm 0.09}$ & 0.48 & $+0.10_{\scriptscriptstyle \pm 0.09}$ & 0.50 & $+0.12_{\scriptscriptstyle \pm 0.10}$ & 0.49 & $+0.08_{\scriptscriptstyle \pm 0.09}$ & \cellcolor{aliceblue}0.49 & \cellcolor{aliceblue}$+0.10_{\scriptscriptstyle \pm 0.09}$ \\
& DE   & 0.49 & $+0.04_{\scriptscriptstyle \pm 0.03}$ & 0.49 & $+0.04_{\scriptscriptstyle \pm 0.05}$ & 0.37 & $+0.09_{\scriptscriptstyle \pm 0.06}$ & 0.67 & $+0.13_{\scriptscriptstyle \pm 0.06}$ & \cellcolor{aliceblue}0.51 & \cellcolor{aliceblue}$+0.08_{\scriptscriptstyle \pm 0.05}$ \\
& EN   & 0.53 & $+0.02_{\scriptscriptstyle \pm 0.03}$ & 0.54 & $+0.03_{\scriptscriptstyle \pm 0.02}$ & 0.37 & $+0.04_{\scriptscriptstyle \pm 0.05}$ & 0.45 & $+0.04_{\scriptscriptstyle \pm 0.06}$ & \cellcolor{aliceblue}0.47 & \cellcolor{aliceblue}$+0.03_{\scriptscriptstyle \pm 0.04}$ \\
& FR   & 0.56 & $+0.07_{\scriptscriptstyle \pm 0.04}$ & 0.49 & $+0.06_{\scriptscriptstyle \pm 0.05}$ & 0.41 & $+0.07_{\scriptscriptstyle \pm 0.06}$ & 0.56 & $+0.08_{\scriptscriptstyle \pm 0.05}$ & \cellcolor{aliceblue}0.51 & \cellcolor{aliceblue}$+0.07_{\scriptscriptstyle \pm 0.05}$ \\
& IT   & 0.58 & $+0.03_{\scriptscriptstyle \pm 0.06}$ & 0.45 & $+0.15_{\scriptscriptstyle \pm 0.06}$ & 0.37 & $+0.05_{\scriptscriptstyle \pm 0.04}$ & 0.52 & $+0.07_{\scriptscriptstyle \pm 0.07}$ & \cellcolor{aliceblue}0.48 & \cellcolor{aliceblue}$+0.08_{\scriptscriptstyle \pm 0.06}$ \\
& JA   & 0.47 & $+0.07_{\scriptscriptstyle \pm 0.05}$ & 0.47 & $+0.04_{\scriptscriptstyle \pm 0.04}$ & 0.43 & $+0.08_{\scriptscriptstyle \pm 0.08}$ & 0.44 & $+0.04_{\scriptscriptstyle \pm 0.03}$ & \cellcolor{aliceblue}0.45 & \cellcolor{aliceblue}$+0.06_{\scriptscriptstyle \pm 0.05}$ \\
& PT   & 0.49 & $+0.07_{\scriptscriptstyle \pm 0.05}$ & 0.50 & $+0.06_{\scriptscriptstyle \pm 0.05}$ & 0.41 & $+0.12_{\scriptscriptstyle \pm 0.06}$ & 0.49 & $+0.08_{\scriptscriptstyle \pm 0.06}$ & \cellcolor{aliceblue}0.47 & \cellcolor{aliceblue}$+0.08_{\scriptscriptstyle \pm 0.06}$ \\
& ZH   & 0.46 & $+0.06_{\scriptscriptstyle \pm 0.04}$ & 0.44 & $+0.05_{\scriptscriptstyle \pm 0.04}$ & 0.46 & $+0.07_{\scriptscriptstyle \pm 0.06}$ & 0.39 & $+0.07_{\scriptscriptstyle \pm 0.04}$ & \cellcolor{aliceblue}0.44 & \cellcolor{aliceblue}$+0.06_{\scriptscriptstyle \pm 0.05}$ \\
\cmidrule(lr){2-12}
\rowcolor{aliceblue}
& \textbf{Avg.} & 0.51 & $+0.06_{\scriptscriptstyle \pm 0.05}$ & 0.48 & $+0.07_{\scriptscriptstyle \pm 0.05}$ & 0.43 & $+0.08_{\scriptscriptstyle \pm 0.06}$ & 0.50 & $+0.07_{\scriptscriptstyle \pm 0.05}$ & 0.48 & $+0.07_{\scriptscriptstyle \pm 0.05}$ \\
\bottomrule
\end{tabular}
}
\vspace{-10pt}
\label{tab:backpatching_final}
\end{table*}

\subsection{The Suppression Effect}



 The ablation results show a substantial shift in modality reliance. In the baseline (Full) state, both models exhibit low audio accuracy, reflecting strong text dominance. However, when the text circuit is ablated ($-\mathcal{C}_T$), audio accuracy increases—for instance, rising from 0.33 to 0.82 in Qwen for the Adjective task. This confirms a direct causal relationship: the text circuit actively suppresses the audio pathway, serving as the important bottleneck that enforces text-dominant decisions. To further scrutinize this cross-modal interaction, we examine the role of the audio circuit by selectively ablating it ($-\mathcal{C}_A$) at positions $\mathcal{P}_A$. Notably, removing the audio circuit drives the already-low accuracy even lower. This further drop indicates that the audio circuit contributes to audio-consistent predictions under conflict, though it is ultimately overwhelmed under conflict conditions.

\subsection{Limitations of Circuit Deletion} 
To explore their collective interaction, we ablate both circuits simultaneously ($-\mathcal{C}_U$). If text dominance were confined entirely to the identified text and audio circuits, ablating both might be expected to move the model toward a more neutral baseline. Instead, we observe asymmetric baselines: Qwen approaches equilibrium, whereas Ultravox retains a degree of text dominance. This discrepancy indicates that text dominance is not confined to the core circuits, but is heavily enforced by the model's background circuits and core language capabilities.

Furthermore, while this asymmetric bias is exposed within a controlled, closed-set classification, relying on disruptive component deletion fundamentally compromises the model's architecture. In practical free-form generation where linguistic coherence is strictly required, such brute-force interventions are typically unviable. This structural limitation indicates that circuit deletion is inadequate as a mitigation strategy, motivating a more precise, non-destructive activation steering approach.


\section{Mitigating Text Dominance via Back-patching}
In this section, we first detail the back-patching approach designed to alleviate textual suppression. We then evaluate its impact on text dominance across an extensive benchmark consisting of two models, eight languages, and four conflict tasks.

\subsection{Back-patching} To counteract this textual suppression, we utilize back-patching~\cite{biran2024hopping, lepori2025racing, same} to transfer acoustic representations across layers and conditions. Instead of destructively turning circuits off, we extract the audio span representations from a late layer and inject them into earlier layers. This design is motivated by the findings in~\cref{subsec:circuit} and~\cref{{subsec:causal}}:~\cref{subsec:circuit} shows that audio representations in the data span converge toward a shared semantic space with text in deeper layers, while~\cref{{subsec:causal}} shows through causal ablation that audio information is not absent but suppressed under conflict. Together, these results suggest that restoring late-layer audio representations to earlier layers may help the audio pathway exert greater influence under conflict. By injecting these acoustic features before the textual bottleneck, we amplify the audio signal and help the model reduce the effect of suppression. Such an intervention indicates that the transferred representation carries causally useful information for mitigating textual dominance.


For each conflict example, our intervention targets the continuous audio token span ($S_{\text{audio}}$) rather than individual isolated tokens. This span-level approach captures acoustic information distributed across multiple tokens, enabling a modality-level causal analysis while reducing computational overhead. We first perform a forward pass through the source condition and record the hidden states for a window of consecutive layers centered at $l_{src}$. Specifically, we cache the activations $\mathbf{h}_{src}^{(l)}$ for $l \in \{l_{src}-i, \dots, l_{src}+i\}$, where the window size $w = 2i + 1$. We then perform a second forward pass under the destination condition. This time, we replace the activations at a corresponding window centered at $l_{dst}$ with our previously recorded source activations. Formally, letting $S_{audio}$ denote the audio token span, the back-patching operation across the layer window is defined as:
\begin{equation}
\begin{aligned}
\mathbf{h}_{dst}^{(l_{dst} + j)}[S_{\text{audio}}]
&\leftarrow
\mathbf{h}_{src}^{(l_{src} + j)}[S_{\text{audio}}] \\
&\forall j \in \{-i, \dots, i\}
\end{aligned}
\end{equation}
In a model with $L$ layers, we systematically explore this transfer by varying the source and destination layers. Because our goal is to explicitly override textual suppression using fully formed acoustic representations, we constrain the search such that $l_{dst} < l_{src}$. We then choose the best-performing valid source-destination configuration on a small subset of the dataset while maintaining the same window size for both source and destination. Detailed analysis of this hyperparameter search is provided in~\cref{appdx:optimal}.



\subsection{Mitigation Results and Bootstrap Variability}
To ensure the reliability of the observed improvements, we evaluate statistical significance using bootstrap resampling with 1,000 iterations.~\cref{tab:backpatching_final} presents the back-patching results in various languages and tasks, reporting the accuracy, the improvement over the baseline ($\Delta$) and the standard deviations between the resamples.

\noindent\textbf{Efficacy of modality restoration.}
By priming the early layers with mature audio representations, our intervention overrides internal text suppression and restores modality balance. This recovery is particularly pronounced in Qwen, yielding an average improvement of $+0.16$ in audio accuracy. The effect is also consistent across multilingual contexts, with especially large gains in Arabic (AR) Adjective Swap ($+0.30_{\pm 0.09}$) and German (DE) Adjective Swap ($+0.31_{\pm 0.09}$). This indicates that even under strong initial text dominance, injecting mature representations can reinforce suppressed acoustic signals. To further examine this effect, we analyze post-intervention shifts in hidden-state L2 norm and attention distributions, which reveal a clear structural prioritization of audio information (detailed in~\cref{appdx:dynamics_backpatching}).
\noindent\textbf{Cross-architectural convergence.}
To contextualize these findings, we compare the intervention effects between Qwen and Ultravox. While Ultravox generally improves across all setups, its absolute gain ($+0.07$) is more tempered than Qwen's ($+0.16$). Crucially, This difference partly reflects their different baseline accuracies: Ultravox starts with a higher average baseline audio accuracy ($0.41$ vs. $0.30$) under this evaluation setup. Despite these disparate starting points, both models ultimately converge to comparable final audio accuracies ($0.46$ for Qwen and $0.48$ for Ultravox) after back-patching. This alignment suggests that the intervention moves both models toward a similar operating point despite their different baseline behaviors. Additional evaluations show positive average gains on three other models and model-dependent improvements with natural voice, while a fixed-path transfer test further verifies the effect without per-setting tuning (detailed in ~\cref{appdx:additional_models,appdx:fixed_path_transfer,appdx:natural_voice_backpatching}).

\section{Conclusion}
In this work, we presented a mechanistic investigation of text dominance in Audio LLMs. By analyzing text-audio conflicts, we identified modality-specific circuits and showed that, although text and audio follow distinct processing pathways, their representations converge toward a shared semantic space in deeper layers. Our causal ablations further demonstrated that acoustic information is not absent under conflict, but instead suppressed by text-dominant processing. These findings motivate back-patching, a training-free intervention that restores late-layer audio representations to earlier layers and improves modality balance under conflict. For future work, we aim to study more precisely where and how audio and text representations become integrated in Audio LLMs.



\section{Limitations}
While our study provides mechanistic insights, it has several limitations. First, our experiments focused on literal speech, without exploring audio-exclusive acoustic attributes such as emotion, prosody, or speaker identity. Evaluating how models integrate these non-verbal representations remains an important direction for future work. Second, while back-patching shows meaningful gains in resolving modality conflicts, the optimal layers for circuit extraction and intervention vary across languages and prompt structures. This underscores the need for a more comprehensive cross-lingual analysis to better generalize internal intervention strategies across diverse language families.

\section{Ethical Considerations}
While our proposed method helps mitigate bias in Audio LLMs, it also carries potential risks of misuse. The ability to manipulate internal mechanisms could be exploited for adversarial attacks, such as maliciously injecting specific biases or suppressing a target modality. 
Our experiments are based on the ALME Benchmark~\cite{alme}. To construct the evaluation data used in this work, we further processed the dataset with Microsoft Edge's online text-to-speech service via \texttt{edge-tts}~\footnote{\url{https://github.com/rany2/edge-tts}}. We summarize the licenses of the datasets, pretrained models, and related assets used in this work in~\cref{sec:asset_licenses}. Therefore, caution is advised when applying these mechanistic interpretation techniques.

\section{Acknowledgements}
This work was supported by Institute of Information \& communications Technology Planning \& Evaluation (IITP) grant funded by the Korea government (MSIT) (RS-2025-02215122, Development and Demonstration of Lightweight AI Model for Smart Homes).


\bibliography{shortstrings, mybibs}

\newpage
\appendix

\label{sec:appendix}
\section*{\centering\LARGE Appendix}
\startcontents[appendixtoc]
\printcontents[appendixtoc]{l}{1}{\setcounter{tocdepth}{2}}
\addtocontents{toc}{\protect\setcounter{tocdepth}{2}}
\definecolor{linkcolor}{HTML}{000000}
\newpage

\definecolor{linkcolor}{HTML}{ED1C24}

\section{Preliminaries}
\label{appdx:circuit_definition}

\subsection{Model Formulation and Task Metric}
Let $\mathcal{M}$ be a transformer-based language model and $x$ be an input token sequence. We represent the computational graph of $\mathcal{M}$ as a set of distinct components (or nodes) $\mathcal{V}$~\cite{conmy2023towards}. For a given input $x$, the activation state of a specific component $v \in \mathcal{V}$ is denoted as $a_v(x)$. To evaluate the model's performance on a specific task, we define a scalar task metric $\mathcal{G}(\mathcal{M}(x))$, such as the logit difference~\cite{wang2022interpretability} between the correct and incorrect target tokens.

\subsection{Circuit Definition}
Building upon the circuit definitions introduced by \cite{same}, we formalize a circuit $c$ as a subset of the model's computational graph $\mathcal{V}$. To precisely locate a component within the transformer architecture, we assign a 3-tuple coordinate $(l, p, i)$ to each node. Thus, a circuit $c \subseteq \mathcal{V}$ is a subset of components indexed by layer, position, and sub-module, i.e., $c \subseteq \{(l, p, i) \in \mathcal{V}\}$. In our experiments, circuits are instantiated by selecting top-ranked components under AP-IG at a chosen sparsity level.
Here, $l$ indicates the layer depth, $p$ denotes the token position in the sequence, and $i$ uniquely identifies the specific sub-module (e.g., an attention head or an MLP neuron).

In our specific task formulation, the functional role of a component heavily depends on its input position. Therefore, we partition the entire circuit $c$ into three distinct functional groups based on their spatial locations: data ($c^D$), query ($c^Q$), and generation ($c^G$) circuits. 
$$c^D = \{(l, p, i) \in c \mid p \in P_D\}$$
$$c^Q = \{(l, p, i) \in c \mid p \in P_Q\}$$
$$c^G = \{(l, p, i) \in c \mid p \in P_G\}$$
We define $P_D, P_Q$, and $P_G$ as mutually exclusive sets of position indices that map to the data context, the query token, and the generation step, respectively. In the main text, $\mathcal{P}_T$ and $\mathcal{P}_A$ denote modality-specific text and audio spans used for causal ablation, whereas $P_D, P_Q, P_G$ partition positions by functional role for circuit analysis. Note that $P_G$ targets the final token position where the model predicts the output.

\subsection{Activation Patching and Circuit Evaluation}
To rigorously evaluate whether a discovered circuit $c$ captures the target behavior, we employ activation patching (causal intervention)~\cite{meng2022rome}. We define two types of inputs: a clean input $x_{clean}$ that elicits the target behavior, and a counterfactual input $x_{cf}$ that serves as a baseline. 

We define a patched model $\mathcal{M}_c(x_{clean}, x_{cf})$ where the activations of components within the circuit $c$ are computed over $x_{clean}$, while the activations of components outside the circuit ($\mathcal{V} \setminus c$) are patched with those from $x_{cf}$. Formally, the activation of component $v$ during the patched forward pass is intervened as:
$$a_v = \begin{cases} a_v(x_{clean}) & \text{if } v \in c \\ a_v(x_{cf}) & \text{if } v \notin c \end{cases}$$
For notational alignment with the main text, $a_v(x_{clean})$ and $a_v(x_{cf})$ are subsequently abbreviated as $a_v^{(clean)}$ and $a_v^{(cf)}$, respectively.

To quantify the faithfulness of the circuit $c$, we measure how much of the clean performance is recovered by the patched model. The faithfulness score $\mathcal{S}(c)$ is computed as:
$$\mathcal{S}(c) = \frac{\mathcal{G}(\mathcal{M}_c(x_{clean}, x_{cf})) - \mathcal{G}(\mathcal{M}(x_{cf}))}{\mathcal{G}(\mathcal{M}(x_{clean})) - \mathcal{G}(\mathcal{M}(x_{cf}))}$$
A score of $\mathcal{S}(c) \approx 1$ indicates that the sub-network $c$ alone is sufficient to perform the task, cleanly isolating the necessary mechanism from the rest of the model.

\section{Implementation Details}
To support the reproducibility of our study, this section provides comprehensive implementation details that supplement the core experimental framework described in the main text. 

\subsection{Dataset Construction and Statistics}
\label{appx:infer_dataset_construction}

To systematically investigate text dominance under multimodal conflict, we extended the ALME benchmark~\cite{alme} to construct a symmetric evaluation setup. While the original benchmark provides pairs of base and counterfactual (CF) text prompts, it only contains natural voice audio for the base text. To bridge this gap, we synthesized Text-to-Speech (TTS) audio for both base and CF text components. To introduce acoustic diversity, we utilized a multilingual neural voice pool with balanced gender distribution (\cref{tab:tts_config}), applying subtle, context-specific adjustments to the speaking rate, time-stretch ratio, and tail padding. 

To satisfy the strict token-length requirements of circuit discovery and evaluation, we leveraged \texttt{gpt-4o-mini}~\cite{hurst2024gpt} to generate alternative candidate sentences that precisely match the required length without altering the core semantic meaning or the ground-truth answer. We applied TTS to these calibrated candidates with only minimal adjustments to speaking rate and padding, strictly discarding any outputs that failed to yield uniform audio token lengths. For causal ablation, we extracted a perfectly aligned text-audio subset that simultaneously satisfies this dual-modality length constraint. Conversely, back-patching experiments utilized randomly selected text-audio pairs without length constraints, as this intervention operates independently of strict token alignment.

\begin{table}[!t]
\centering
\small
\caption{\textbf{Summary of TTS dataset configurations.} The table details the pool size and gender distribution across different languages.}
\label{tab:tts_config}
\begin{tabular}{c c c}
\toprule
\textbf{Lang.} & \textbf{Pool Size} & \textbf{Gender Mix} \\ 
\midrule
AR & 4 & 2F / 2M \\ 
DE & 6 & 3F / 3M \\ 
EN & 7 & 4F / 3M \\
FR & 6 & 3F / 3M \\ 
IT & 4 & 2F / 2M \\
JA & 2 & 1F / 1M \\
PT & 4 & 2F / 2M \\ 
ZH & 8 & 4F / 4M \\ 
\bottomrule
\end{tabular}
\end{table}

\begin{table}[!t]
\centering
\small
\caption{\textbf{Total number of prompts used for circuit discovery in the EN dataset.} The "Text" and "Audio" columns denote the number of length-matched samples strictly curated to isolate the text and audio pathways, respectively.}
\label{tab:circuit_quantity}
\setlength{\tabcolsep}{3pt} 
\begin{tabular}{c c c c c c}
\toprule
& & \multicolumn{2}{c}{\textbf{Qwen}} & \multicolumn{2}{c}{\textbf{Ultravox}} \\
\cmidrule{3-4} \cmidrule{5-6}
\multirow{-2}{*}{\textbf{Lang.}} & \multirow{-2}{*}{\textbf{Task}} & \textbf{Text} & \textbf{Audio} & \textbf{Text} & \textbf{Audio} \\
\midrule
& Adj. Swap  & 353 & 322 & 352 & 255 \\
& Neg. Swap  & 258 & 248 & 258 & 246 \\
& Num. Swap  & 431 & 330 & 430 & 367 \\
\multirow{-4}{*}{EN} & Time Swap  & 332 & 291 & 335 & 306 \\
\bottomrule
\end{tabular}
\end{table}

\begin{table}[!t]
\centering
\small
\caption{\textbf{Total number of prompts used in the causal ablation experiment.} The table lists the final counts evaluated on the EN dataset for Qwen and Ultravox.}
\label{tab:causal_quantity} 
\begin{tabular}{l c c}
\toprule
\textbf{Task} & \textbf{Qwen} & \textbf{Ultra.} \\
\midrule
Adj. Swap & 322 & 255 \\
Neg. Swap & 248 & 246 \\
Num. Swap & 330 & 367 \\
Time Swap & 291 & 306 \\
\bottomrule
\end{tabular}
\end{table}


We configured the evaluation and analysis datasets with varying sample scales tailored to each experiment. Specifically, behavioral inference~(\S\ref{subsec:dataset}) was conducted using 250 samples per task and language. For circuit analysis~(\S\ref{subsec:circuit}), the total data volume is outlined in \cref{tab:circuit_quantity}. The sample breakdown for causal ablation~(\S\ref{subsec:causal}) is documented in \cref{tab:causal_quantity}. For the back-patching intervention experiments, we evaluated performance using 100 samples per task and language.


\subsection{Experimental Setup for Circuit Discovery and Evaluation}
\label{appdx:circuit_setup}
Specifically, our curated dataset was partitioned into a 75\% split for circuit discovery and a 25\% split for evaluation. To ensure that the models could fully determine the correct answer at the very first generated token, we constrained the output space to a multiple-choice format, restricting the generation to either "A" or "B." Furthermore, to eliminate positional bias and ensure fairness, the label assignments for the choice options were completely randomized for each sample.

\begin{figure}[!t]
    \centering
    \includegraphics[width=\linewidth]{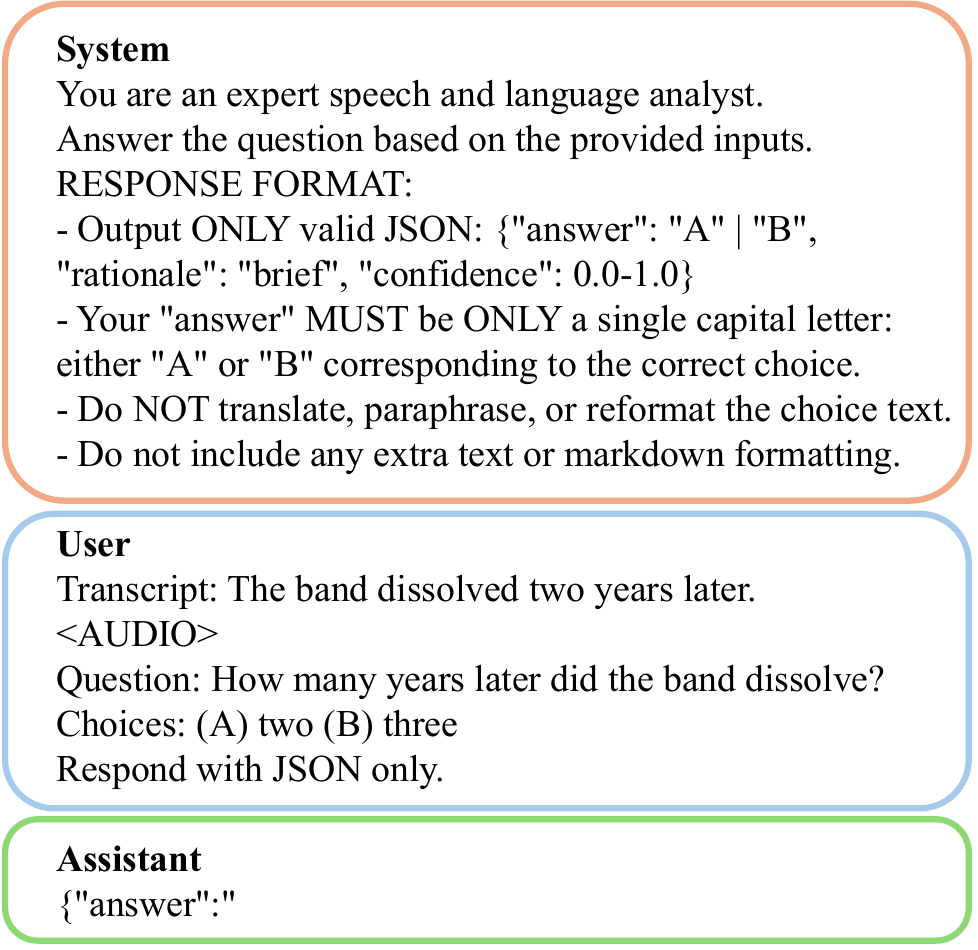}
   \caption{\textbf{Prompt template example for the English Number Swap task applied to Qwen}, illustrating a multimodal setup where text data precedes audio data. The prompt design is adapted from ALME~\cite{alme}, and a structurally similar template is employed for the Ultravox model.}
    \label{fig:prompt_example_qwen}
\end{figure}

For Qwen2-Audio, this constraint was successfully enforced by structuring the prompt template as illustrated in~\cref{fig:prompt_example_qwen}. This template was similarly adapted for Ultravox to guarantee consistent multiple-choice generation across both architectures. All experiments, including the intensive circuit discovery phase, were conducted on a single compute node equipped with 8 NVIDIA RTX A5000 GPUs (24GB VRAM each).

\section{Evidence of Text Dominance}
In this section, we provide the detailed behavioral results that complement the averaged metrics presented in the main paper (\cref{tab:motivation_subscript}).

\subsection{Free Generation with Natural Voice}
To evaluate the models under realistic, unconstrained inference conditions, \cref{tab:behavioral_results_natural} details the performance of Qwen2-Audio and Ultravox across eight languages under the free generation setting. The aligned baselines show high accuracy across languages, suggesting that these Audio LLMs retain strong performance when modalities agree. In contrast, the conflict views reveal a persistent text bias. Notably, the dominance margin ($\Delta_{T-A}$) remains predominantly positive across the multilingual spectrum. However, the magnitude of this bias varies depending on the specific conflict direction (e.g., $A_{\text{base}} \Rightarrow T_{\text{CF}}$ versus $T_{\text{CF}} \Rightarrow A_{\text{base}}$).

\subsection{Limited Labels Setting}
The averaged results in the main text (\cref{tab:motivation_subscript}) were conducted under a limited labels setting to isolate and quantify the models' modality preferences, preventing the natural variance of open-ended generation from obscuring the underlying bias. Building on this, \cref{tab:behavioral_results_tts} extends our granular analysis to this limited labels setting using synthetic TTS data, covering four distinct conflict configurations. Consistent with the natural voice results, we observe a prominent text-affinity across the majority of languages and scenarios for both models. These detailed breakdowns suggest that the modality bias discussed in the main text appears across different languages rather than being an artifact of any single linguistic subset.

\begin{table*}[p]
\centering
\small
\caption{\textbf{Behavioral accuracy under natural voice and free generation settings across eight languages.} Serving as the empirical motivation for our study, these results show text-dominant behavior in both Qwen and Ultravox. For the conflict view, we report text-affinity accuracy and the corresponding dominance margin ($\Delta_{T-A} = \text{text acc.} - \text{audio acc.}$), where bold values highlight a prominent text bias ($\Delta_{T-A} > 0.10$). Aligned Avg. represents the baseline performance without modality conflict.}
\label{tab:behavioral_results_natural}
\begin{tabular}{lcccccccccc}
\toprule
 & \multicolumn{4}{c}{\textbf{Qwen2-Audio}} & \phantom{a} & \multicolumn{4}{c}{\textbf{Ultravox}} \\
\cmidrule{2-5} \cmidrule{7-10}
 & \multicolumn{2}{c}{Conflict View (Text / $\Delta_{T-A}$)} & \multicolumn{2}{c}{Average} & & \multicolumn{2}{c}{Conflict View (Text / $\Delta_{T-A}$)} & \multicolumn{2}{c}{Average} \\
\cmidrule{2-3} \cmidrule{4-5} \cmidrule{7-8} \cmidrule{9-10}
\textbf{Lang.} & $A_{\text{base}} \rightarrow T_{\text{CF}}$ & $T_{\text{CF}} \rightarrow A_{\text{base}}$ & Conflict & Aligned & & $A_{\text{base}} \rightarrow T_{\text{CF}}$ & $T_{\text{CF}} \rightarrow A_{\text{base}}$ & Conflict & Aligned \\
\midrule
AR & 0.755 / \textbf{+0.511} & 0.613 / \textbf{+0.226} & 0.684 & 0.869 & & 0.592 / \textbf{+0.184} & 0.640 / \textbf{+0.280} & 0.616 & 0.922 \\
DE & 0.731 / \textbf{+0.462} & 0.463 / -0.073          & 0.597 & 0.947 & & 0.528 / +0.056          & 0.636 / \textbf{+0.272} & 0.582 & 0.944 \\
EN & 0.717 / \textbf{+0.433} & 0.367 / -0.266          & 0.542 & 0.928 & & 0.524 / +0.048          & 0.504 / +0.008          & 0.514 & 0.944 \\
FR & 0.715 / \textbf{+0.431} & 0.459 / -0.081          & 0.587 & 0.929 & & 0.596 / \textbf{+0.192} & 0.560 / \textbf{+0.120} & 0.578 & 0.900 \\
IT & 0.696 / \textbf{+0.393} & 0.375 / -0.250          & 0.535 & 0.955 & & 0.540 / +0.080          & 0.516 / +0.032          & 0.528 & 0.958 \\
JA & 0.643 / \textbf{+0.285} & 0.488 / -0.024          & 0.565 & 0.885 & & 0.556 / \textbf{+0.112} & 0.660 / \textbf{+0.320} & 0.608 & 0.896 \\
PT & 0.704 / \textbf{+0.407} & 0.458 / -0.084          & 0.581 & 0.912 & & 0.556 / \textbf{+0.112} & 0.552 / \textbf{+0.104} & 0.554 & 0.928 \\
ZH & 0.688 / \textbf{+0.376} & 0.423 / -0.153          & 0.555 & 0.954 & & 0.552 / \textbf{+0.104} & 0.704 / \textbf{+0.408} & 0.628 & 0.940 \\
\bottomrule
\end{tabular}
\end{table*}

\begin{table*}[p]
\centering
\small
\caption{\textbf{Behavior accuracy under the limited labels setting with synthetic TTS data across eight languages.} Complementing the natural voice results, this table shows that text-dominant behavior also appears under closed-set evaluation with synthetic audio. We report the text-affinity accuracy and the dominance margin ($\Delta_{T-A} = \text{text acc.} - \text{audio acc.}$), with bold values indicating a prominent text bias ($\Delta_{T-A} > 0.10$).}
\label{tab:behavioral_results_tts}
\begin{tabular}{ccccccc}
\toprule
\textbf{Language} & $A_{\text{base}} \rightarrow T_{\text{CF}}$ & $A_{\text{CF}} \rightarrow T_{\text{base}}$ & $T_{\text{base}} \rightarrow A_{\text{CF}}$ & $T_{\text{CF}} \rightarrow $A$_{\text{base}}$ & \textbf{Conflict Avg.} & \textbf{Aligned Avg.} \\
\midrule
\multicolumn{7}{c}{\textbf{Qwen}} \\
\midrule
AR & 0.708 / \textbf{+0.416} & 0.760 / \textbf{+0.520} & 0.672 / \textbf{+0.344} & 0.588 / \textbf{+0.176} & 0.682 / \textbf{+0.364} & 0.858 \\
DE & 0.740 / \textbf{+0.480} & 0.708 / \textbf{+0.416} & 0.484 / -0.032          & 0.476 / -0.048          & 0.602 / \textbf{+0.204} & 0.912 \\
EN & 0.740 / \textbf{+0.480} & 0.664 / \textbf{+0.328} & 0.432 / -0.136          & 0.428 / -0.144          & 0.566 / \textbf{+0.132} & 0.906 \\
FR & 0.716 / \textbf{+0.432} & 0.684 / \textbf{+0.368} & 0.412 / -0.176          & 0.416 / -0.168          & 0.557 / \textbf{+0.114} & 0.922 \\
IT & 0.684 / \textbf{+0.368} & 0.772 / \textbf{+0.544} & 0.512 / +0.024          & 0.336 / -0.328          & 0.576 / \textbf{+0.152} & 0.958 \\
JA & 0.660 / \textbf{+0.320} & 0.636 / \textbf{+0.272} & 0.544 / +0.088          & 0.572 / \textbf{+0.144} & 0.603 / \textbf{+0.206} & 0.876 \\
PT & 0.720 / \textbf{+0.440} & 0.612 / \textbf{+0.224} & 0.424 / -0.152          & 0.516 / +0.032          & 0.568 / \textbf{+0.136} & 0.896 \\
ZH & 0.688 / \textbf{+0.376} & 0.728 / \textbf{+0.456} & 0.500 / +0.000          & 0.452 / -0.096          & 0.592 / \textbf{+0.184} & 0.924 \\
\midrule
\multicolumn{7}{c}{\textbf{Ultravox}} \\
\midrule
AR & 0.564 / \textbf{+0.128} & 0.664 / \textbf{+0.328} & 0.692 / \textbf{+0.384} & 0.632 / \textbf{+0.264} & 0.638 / \textbf{+0.276} & 0.904 \\
DE & 0.528 / +0.056          & 0.608 / \textbf{+0.216} & 0.632 / \textbf{+0.264} & 0.588 / \textbf{+0.176} & 0.589 / \textbf{+0.178} & 0.938 \\
EN & 0.528 / +0.056          & 0.568 / \textbf{+0.136} & 0.640 / \textbf{+0.280} & 0.552 / \textbf{+0.104} & 0.572 / \textbf{+0.144} & 0.950 \\
FR & 0.560 / \textbf{+0.120} & 0.568 / \textbf{+0.136} & 0.604 / \textbf{+0.208} & 0.596 / \textbf{+0.192} & 0.582 / \textbf{+0.164} & 0.946 \\
IT & 0.576 / \textbf{+0.152} & 0.656 / \textbf{+0.312} & 0.676 / \textbf{+0.352} & 0.528 / +0.056          & 0.609 / \textbf{+0.218} & 0.966 \\
JA & 0.568 / \textbf{+0.136} & 0.632 / \textbf{+0.264} & 0.728 / \textbf{+0.456} & 0.700 / \textbf{+0.400} & 0.657 / \textbf{+0.314} & 0.884 \\
PT & 0.580 / \textbf{+0.160} & 0.620 / \textbf{+0.240} & 0.568 / \textbf{+0.136} & 0.592 / \textbf{+0.184} & 0.590 / \textbf{+0.180} & 0.888 \\
ZH & 0.576 / \textbf{+0.152} & 0.672 / \textbf{+0.344} & 0.748 / \textbf{+0.496} & 0.692 / \textbf{+0.384} & 0.672 / \textbf{+0.344} & 0.922 \\
\bottomrule
\end{tabular}
\end{table*}

\begin{figure*}[tp]
    \centering
    \includegraphics[width=\textwidth]{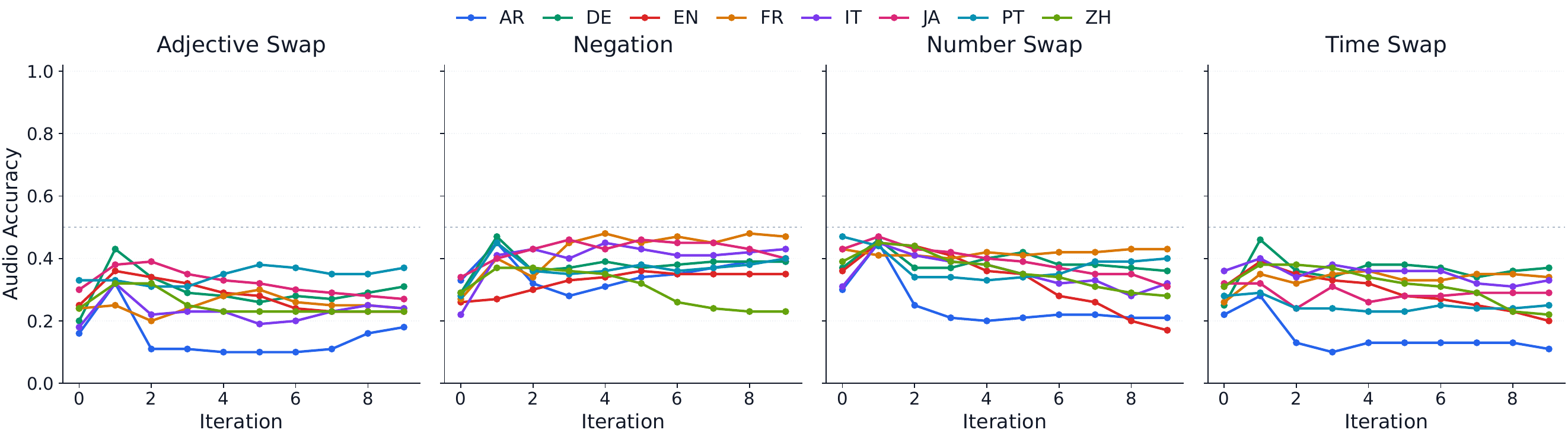}
    \caption{\textbf{Effect of iterative back-patching on audio accuracy.} The results illustrate that applying the intervention iteratively does not guarantee continuous improvements. In most cases, a single back-patching iteration yields the maximum performance gain, while subsequent iterations result in diminishing returns or slight regressions.}    \label{fig:iterative_patching}
\end{figure*}

\begin{table*}[htbp]
\centering
\caption{\textbf{Best back-patching configurations across models, languages, and flip types.} Each cell presents the source layer ($l_{src}$), destination layer ($l_{dst}$), and layer window size (in parentheses) that lead to the highest back-patching accuracy, formatted as $l_{src} \rightarrow l_{dst}$ (window size).}
\label{tab:backpatching_best_config}
\resizebox{\textwidth}{!}{%
\begin{tabular}{l l c c c c c c c c}
\toprule
\textbf{Model} & \textbf{Flip Type} & \textbf{AR} & \textbf{DE} & \textbf{EN} & \textbf{FR} & \textbf{IT} & \textbf{JA} & \textbf{PT} & \textbf{ZH} \\
\midrule
\multirow{4}{*}{Qwen} 
& Adj. Swap   & 31 $\rightarrow$ 1 (3)  & 31 $\rightarrow$ 2 (3)  & 15 $\rightarrow$ 4 (3)  & 31 $\rightarrow$ 2 (3)  & 31 $\rightarrow$ 1 (1)  & 11 $\rightarrow$ 6 (1)  & 31 $\rightarrow$ 2 (5)  & 17 $\rightarrow$ 0 (1)  \\
& Neg. Swap   & 31 $\rightarrow$ 0 (1)  & 31 $\rightarrow$ 1 (1)  & 31 $\rightarrow$ 6 (3)  & 31 $\rightarrow$ 3 (3)  & 31 $\rightarrow$ 2 (1)  & 31 $\rightarrow$ 1 (1)  & 31 $\rightarrow$ 0 (3)  & 13 $\rightarrow$ 5 (1)  \\
& Num. Swap   & 31 $\rightarrow$ 0 (3)  & 31 $\rightarrow$ 0 (3)  & 12 $\rightarrow$ 4 (1)  & 31 $\rightarrow$ 0 (1)  & 31 $\rightarrow$ 2 (1)  & 11 $\rightarrow$ 4 (3)  & 31 $\rightarrow$ 2 (1)  & 11 $\rightarrow$ 4 (5)  \\
& Time Swap   & 31 $\rightarrow$ 0 (1)  & 31 $\rightarrow$ 2 (3)  & 13 $\rightarrow$ 4 (3)  & 31 $\rightarrow$ 1 (1)  & 31 $\rightarrow$ 2 (3)  & 31 $\rightarrow$ 0 (1)  & 31 $\rightarrow$ 1 (1)  & 12 $\rightarrow$ 3 (5)  \\
\midrule
\multirow{4}{*}{Ultravox} 
& Adj. Swap   & 31 $\rightarrow$ 0 (1)  & 22 $\rightarrow$ 10 (1) & 24 $\rightarrow$ 13 (1) & 29 $\rightarrow$ 12 (1) & 26 $\rightarrow$ 12 (1) & 30 $\rightarrow$ 5 (1)  & 29 $\rightarrow$ 12 (1) & 29 $\rightarrow$ 11 (3) \\
& Neg. Swap   & 31 $\rightarrow$ 1 (3)  & 30 $\rightarrow$ 14 (5) & 10 $\rightarrow$ 9 (3)  & 31 $\rightarrow$ 13 (3) & 27 $\rightarrow$ 9 (5)  & 26 $\rightarrow$ 11 (1) & 31 $\rightarrow$ 13 (5) & 31 $\rightarrow$ 14 (5) \\
& Num. Swap   & 31 $\rightarrow$ 1 (1)  & 31 $\rightarrow$ 0 (1)  & 28 $\rightarrow$ 11 (5) & 29 $\rightarrow$ 8 (3)  & 29 $\rightarrow$ 12 (1) & 31 $\rightarrow$ 6 (1)  & 31 $\rightarrow$ 10 (5) & 30 $\rightarrow$ 9 (1)  \\
& Time Swap   & 31 $\rightarrow$ 1 (1)  & 30 $\rightarrow$ 12 (3) & 29 $\rightarrow$ 3 (5)  & 30 $\rightarrow$ 12 (1) & 29 $\rightarrow$ 8 (3)  & 23 $\rightarrow$ 8 (1)  & 30 $\rightarrow$ 9 (1)  & 27 $\rightarrow$ 11 (1)  \\
\bottomrule
\end{tabular}%
}
\end{table*}

\begin{table*}[tp]
\centering
\small
\caption{\textbf{Back-patching performance under counterfactual audio conflicts using \textit{natural voice}.} Due to the unidirectional availability of natural voice samples, results are evaluated exclusively under the setup where a counterfactual cue is presented in the audio alongside a literal text prompt. We report the patched audio accuracy and the improvement ($\Delta$, relative to the unpatched baseline). In this controlled binary setting, an audio accuracy of 0.5 serves as a balanced reference point rather than a general deployment objective.}
\scalebox{0.98}{
\begin{tabular}{ll
c@{\hspace{2pt}}c@{\hspace{8pt}}
c@{\hspace{2pt}}c@{\hspace{8pt}}
c@{\hspace{2pt}}c@{\hspace{8pt}}
c@{\hspace{2pt}}c@{\hspace{8pt}}
c@{\hspace{2pt}}c}
\toprule
\multirow{2}{*}{\textbf{Model}}
& \multirow{2}{*}{\textbf{Lang.}}
& \multicolumn{2}{c}{\textbf{Adj. Swap}}
& \multicolumn{2}{c}{\textbf{Neg. Swap}}
& \multicolumn{2}{c}{\textbf{Num. Swap}}
& \multicolumn{2}{c}{\textbf{Time Swap}}
& \multicolumn{2}{c}{\textbf{Avg.}} \\
\cmidrule(lr){3-4} \cmidrule(lr){5-6} \cmidrule(lr){7-8} \cmidrule(lr){9-10} \cmidrule(lr){11-12}
& & Patched & $\Delta$ & Patched & $\Delta$ & Patched & $\Delta$ & Patched & $\Delta$ & Patched & $\Delta$ \\
\midrule
\multirow{9}{*}{Qwen}
& AR   & 0.37 & $+0.21$ & 0.45 & $+0.19$ & 0.42 & $+0.22$ & 0.39 & $+0.17$ & 0.41 & $+0.20$ \\
& DE   & 0.45 & $+0.21$ & 0.52 & $+0.27$ & 0.47 & $+0.18$ & 0.43 & $+0.21$ & 0.47 & $+0.22$ \\
& EN   & 0.32 & $+0.10$ & 0.34 & $-0.04$ & 0.32 & $+0.07$ & 0.36 & $+0.05$ & 0.34 & $+0.05$ \\
& FR   & 0.36 & $+0.17$ & 0.40 & $+0.09$ & 0.45 & $+0.19$ & 0.41 & $+0.21$ & 0.41 & $+0.17$ \\
& IT   & 0.45 & $+0.14$ & 0.50 & $+0.18$ & 0.49 & $+0.22$ & 0.42 & $+0.12$ & 0.47 & $+0.17$ \\
& JA   & 0.30 & $+0.05$ & 0.52 & $+0.10$ & 0.33 & $+0.00$ & 0.34 & $+0.03$ & 0.37 & $+0.05$ \\
& PT   & 0.35 & $+0.05$ & 0.43 & $+0.11$ & 0.45 & $+0.23$ & 0.40 & $+0.12$ & 0.41 & $+0.13$ \\
& ZH   & 0.32 & $+0.04$ & 0.43 & $+0.08$ & 0.39 & $+0.02$ & 0.36 & $+0.04$ & 0.38 & $+0.05$ \\
\cmidrule(lr){2-12}
& Avg. & 0.37 & $+0.12$ & 0.45 & $+0.13$ & 0.42 & $+0.14$ & 0.39 & $+0.12$ & 0.41 & $+0.13$ \\
\midrule
\multirow{9}{*}{Ultra.}
& AR   & 0.47 & $-0.02$ & 0.39 & $-0.03$ & 0.46 & $+0.12$ & 0.45 & $+0.01$ & 0.44 & $+0.02$ \\
& DE   & 0.59 & $+0.03$ & 0.45 & $-0.03$ & 0.34 & $+0.06$ & 0.52 & $+0.02$ & 0.48 & $+0.02$ \\
& EN   & 0.55 & $-0.01$ & 0.57 & $+0.02$ & 0.38 & $+0.08$ & 0.51 & $+0.01$ & 0.50 & $+0.03$ \\
& FR   & 0.51 & $+0.00$ & 0.41 & $-0.02$ & 0.31 & $+0.04$ & 0.46 & $+0.01$ & 0.42 & $+0.01$ \\
& IT   & 0.58 & $+0.02$ & 0.50 & $+0.06$ & 0.33 & $+0.02$ & 0.48 & $-0.01$ & 0.47 & $+0.02$ \\
& JA   & 0.47 & $-0.00$ & 0.49 & $+0.06$ & 0.38 & $+0.03$ & 0.43 & $+0.03$ & 0.44 & $+0.03$ \\
& PT   & 0.52 & $+0.02$ & 0.46 & $-0.05$ & 0.18 & $+0.05$ & 0.49 & $-0.01$ & 0.41 & $+0.00$ \\
& ZH   & 0.53 & $+0.05$ & 0.46 & $+0.01$ & 0.41 & $+0.06$ & 0.38 & $+0.02$ & 0.45 & $+0.04$ \\
\cmidrule(lr){2-12}
& Avg. & 0.53 & $+0.01$ & 0.47 & $+0.01$ & 0.36 & $+0.06$ & 0.47 & $+0.01$ & 0.46 & $+0.02$ \\
\bottomrule
\end{tabular}
}
\label{tab:backpatching_natual_voice}
\end{table*}

\begin{table*}[tp]
\centering
\small
\caption{
\textbf{Back-patching results evaluated under \textit{aligned} modality conditions.} For each task, we report the patched audio accuracy and the bootstrap-estimated improvement ($\Delta$) averaged across two complementary aligned setups. When the modalities are already aligned, back-patching yields marginal performance deltas ($\Delta \approx 0.00$) across all models and tasks.
}
\scalebox{0.94}{
\begin{tabular}{ll
c@{\hspace{2pt}}c@{\hspace{8pt}}
c@{\hspace{2pt}}c@{\hspace{8pt}}
c@{\hspace{2pt}}c@{\hspace{8pt}}
c@{\hspace{2pt}}c@{\hspace{8pt}}
c@{\hspace{2pt}}c}
\toprule
\multirow{2}{*}{\textbf{Model}}
& \multirow{2}{*}{\textbf{Lang.}}
& \multicolumn{2}{c}{\textbf{Adj. Swap}}
& \multicolumn{2}{c}{\textbf{Neg. Swap}}
& \multicolumn{2}{c}{\textbf{Num. Swap}}
& \multicolumn{2}{c}{\textbf{Time Swap}}
& \multicolumn{2}{c}{\textbf{Avg.}} \\
\cmidrule(lr){3-4} \cmidrule(lr){5-6} \cmidrule(lr){7-8} \cmidrule(lr){9-10} \cmidrule(lr){11-12}
& & Patched & $\Delta$ & Patched & $\Delta$ & Patched & $\Delta$ & Patched & $\Delta$ & Patched & $\Delta$ \\
\midrule
\multirow{9}{*}{Qwen}
& AR   & 0.95 & $+0.02$ & 0.85 & $-0.01$ & 0.91 & $+0.03$ & 0.87 & $-0.01$ & 0.89 & $+0.01$ \\
& DE   & 0.96 & $-0.00$ & 0.85 & $+0.02$ & 0.95 & $-0.00$ & 0.97 & $-0.01$ & 0.93 & $+0.00$ \\
& EN   & 0.95 & $-0.00$ & 0.87 & $-0.00$ & 0.95 & $+0.00$ & 0.95 & $+0.01$ & 0.93 & $+0.00$ \\
& FR   & 0.94 & $-0.00$ & 0.89 & $+0.01$ & 0.92 & $-0.01$ & 0.96 & $-0.00$ & 0.93 & $-0.00$ \\
& IT   & 0.97 & $+0.01$ & 0.88 & $-0.01$ & 0.94 & $-0.01$ & 0.97 & $-0.01$ & 0.94 & $-0.01$ \\
& JA   & 0.95 & $+0.00$ & 0.84 & $+0.01$ & 0.90 & $+0.01$ & 0.95 & $-0.00$ & 0.91 & $+0.01$ \\
& PT   & 0.94 & $+0.01$ & 0.89 & $+0.00$ & 0.88 & $-0.03$ & 0.94 & $-0.01$ & 0.91 & $-0.01$ \\
& ZH   & 0.95 & $+0.00$ & 0.95 & $+0.02$ & 0.91 & $+0.01$ & 0.94 & $-0.00$ & 0.94 & $+0.01$ \\
\cmidrule(lr){2-12}
& Avg. & 0.95 & $+0.00$ & 0.87 & $+0.01$ & 0.92 & $+0.00$ & 0.94 & $-0.01$ & 0.92 & $+0.00$ \\
\bottomrule
\end{tabular}
}
\label{tab:backpatching_aligned}
\end{table*}

\section{Expanded Analysis of Back-Patching}
\label{appdx:expanded_backpatching}

In this section, we provide extended experimental results for our causal back-patching interventions. These results include hyperparameter configurations, validation on natural human speech, non-destructive sanity checks on aligned data, and additional back-patching heatmap visualization.

\subsection{Additional Model Evaluation}
\label{appdx:additional_models}
To examine whether the intervention behavior extends beyond the two models used in the main analysis, we additionally evaluate back-patching on SALMONN, Audio Flamingo 3, and Qwen3-Omni-30B. SALMONN and Audio Flamingo 3 are evaluated across the full eight-language setting, while Qwen3-Omni-30B is included as an English-only supplementary check due to its higher inference cost. As shown in \cref{tab:appdx_additional_models}, these additional evaluations show positive average improvements in audio accuracy, supporting the broader applicability of the observed intervention effect while preserving the model-dependent interpretation discussed in the main text.

\begin{table*}[h]
\centering
\small
\caption{\textbf{Back-patching on additional Audio LLMs.} We report patched audio accuracy and the improvement ($\Delta$) over the unpatched baseline. SALMONN and Audio Flamingo 3 are evaluated across all eight languages, while Qwen3-Omni-30B is an English-only supplementary check.}
\scalebox{0.94}{
\begin{tabular}{ll c@{\hspace{2pt}}c@{\hspace{8pt}} c@{\hspace{2pt}}c@{\hspace{8pt}} c@{\hspace{2pt}}c@{\hspace{8pt}} c@{\hspace{2pt}}c@{\hspace{8pt}} c@{\hspace{2pt}}c}
\toprule
\multirow{2}{*}{\textbf{Model}}
& \multirow{2}{*}{\textbf{Lang.}}
& \multicolumn{2}{c}{\textbf{Adj. Swap}}
& \multicolumn{2}{c}{\textbf{Neg. Swap}}
& \multicolumn{2}{c}{\textbf{Num. Swap}}
& \multicolumn{2}{c}{\textbf{Time Swap}}
& \multicolumn{2}{c}{\textbf{Avg.}}\\
\cmidrule(lr){3-4} \cmidrule(lr){5-6} \cmidrule(lr){7-8} \cmidrule(lr){9-10} \cmidrule(lr){11-12}
& & Patched & $\Delta$ & Patched & $\Delta$ & Patched & $\Delta$ & Patched & $\Delta$ & Patched & $\Delta$\\
\midrule

\multirow{9}{*}{SALMONN}
& AR & 0.60 & $+0.18$ & 0.57 & $+0.04$ & 0.62 & $+0.12$ & 0.62 & $+0.00$ & \cellcolor{aliceblue}0.60 & \cellcolor{aliceblue}$+0.08$ \\
& DE & 0.62 & $+0.06$ & 0.59 & $+0.07$ & 0.58 & $+0.04$ & 0.64 & $+0.10$ & \cellcolor{aliceblue}0.61 & \cellcolor{aliceblue}$+0.07$ \\
& EN & 0.60 & $+0.06$ & 0.46 & $+0.08$ & 0.58 & $+0.08$ & 0.58 & $+0.12$ & \cellcolor{aliceblue}0.55 & \cellcolor{aliceblue}$+0.08$ \\
& FR & 0.52 & $+0.08$ & 0.56 & $+0.13$ & 0.62 & $+0.06$ & 0.58 & $+0.14$ & \cellcolor{aliceblue}0.57 & \cellcolor{aliceblue}$+0.10$ \\
& IT & 0.60 & $+0.00$ & 0.52 & $+0.02$ & 0.62 & $+0.00$ & 0.58 & $+0.04$ & \cellcolor{aliceblue}0.58 & \cellcolor{aliceblue}$+0.01$ \\
& JA & 0.66 & $+0.08$ & 0.52 & $+0.09$ & 0.54 & $+0.10$ & 0.64 & $+0.08$ & \cellcolor{aliceblue}0.59 & \cellcolor{aliceblue}$+0.09$ \\
& PT & 0.60 & $+0.14$ & 0.60 & $+0.14$ & 0.56 & $+0.04$ & 0.56 & $+0.02$ & \cellcolor{aliceblue}0.58 & \cellcolor{aliceblue}$+0.09$ \\
& ZH & 0.46 & $+0.06$ & 0.59 & $+0.12$ & 0.62 & $+0.20$ & 0.58 & $+0.08$ & \cellcolor{aliceblue}0.56 & \cellcolor{aliceblue}$+0.12$ \\
\cmidrule(lr){2-12}
\rowcolor{aliceblue}
& \textbf{Avg.} & 0.58 & $+0.08$ & 0.55 & $+0.09$ & 0.59 & $+0.08$ & 0.60 & $+0.07$ & 0.58 & $+0.08$ \\

\midrule

\multirow{9}{*}{Audio Flamingo 3}
& AR & 0.49 & $+0.14$ & 0.51 & $+0.19$ & 0.50 & $+0.12$ & 0.49 & $+0.06$ & \cellcolor{aliceblue}0.50 & \cellcolor{aliceblue}$+0.14$ \\
& DE & 0.51 & $+0.07$ & 0.61 & $+0.07$ & 0.52 & $+0.05$ & 0.50 & $+0.01$ & \cellcolor{aliceblue}0.55 & \cellcolor{aliceblue}$+0.05$ \\
& EN & 0.50 & $+0.19$ & 0.54 & $+0.15$ & 0.50 & $+0.23$ & 0.51 & $+0.10$ & \cellcolor{aliceblue}0.52 & \cellcolor{aliceblue}$+0.16$ \\
& FR & 0.49 & $+0.17$ & 0.57 & $+0.16$ & 0.53 & $+0.08$ & 0.50 & $+0.04$ & \cellcolor{aliceblue}0.53 & \cellcolor{aliceblue}$+0.12$ \\
& IT & 0.50 & $+0.06$ & 0.53 & $+0.13$ & 0.50 & $+0.11$ & 0.50 & $+0.02$ & \cellcolor{aliceblue}0.51 & \cellcolor{aliceblue}$+0.09$ \\
& JA & 0.50 & $+0.06$ & 0.54 & $+0.03$ & 0.60 & $+0.11$ & 0.48 & $+0.01$ & \cellcolor{aliceblue}0.53 & \cellcolor{aliceblue}$+0.05$ \\
& PT & 0.50 & $+0.06$ & 0.58 & $+0.16$ & 0.66 & $+0.16$ & 0.50 & $+0.00$ & \cellcolor{aliceblue}0.56 & \cellcolor{aliceblue}$+0.11$ \\
& ZH & 0.48 & $+0.07$ & 0.51 & $-0.02$ & 0.52 & $+0.04$ & 0.51 & $+0.00$ & \cellcolor{aliceblue}0.51 & \cellcolor{aliceblue}$+0.01$ \\
\cmidrule(lr){2-12}
\rowcolor{aliceblue}
& \textbf{Avg.} & 0.50 & $+0.10$ & 0.55 & $+0.11$ & 0.54 & $+0.11$ & 0.50 & $+0.03$ & 0.53 & $+0.09$ \\

\midrule

Qwen3-Omni-30B
& EN & 0.70	& $+0.14$	& 0.58	& $+0.13$	& 0.66	& $+0.18$	& 0.70	& $+0.21$	& 0.64	& $+0.16$ \\

\bottomrule
\end{tabular}
}
\label{tab:appdx_additional_models}
\end{table*}

\subsection{Fixed-Path Transfer Intervention}
\label{appdx:fixed_path_transfer}
The main back-patching experiments select source and destination layers on a small validation subset for each setting. To test whether a single intervention path can transfer across settings, we apply the fixed Qwen2-Audio path $31 \rightarrow 3$ to all languages and conflict types without per-setting tuning. As shown in \cref{tab:qwen_fixed_patch_31_3}, this fixed-path setting still improves average audio accuracy, although the gains vary across languages and task types.

\begin{table*}[t]
\centering
\small
\caption{\textbf{Fixed-path transfer evaluation for back-patching.} We apply the same Qwen2-Audio intervention path ($31 \rightarrow 3$) across all languages and conflict types without per-setting tuning. Each cell reports patched audio accuracy, with the improvement over the unpatched baseline in parentheses.}
\resizebox{0.72\textwidth}{!}{
\begin{tabular}{lccccc}
\toprule
\textbf{Lang.} 
& \textbf{Adj. Swap} 
& \textbf{Neg. Swap} 
& \textbf{Num. Swap} 
& \textbf{Time Swap} 
& \textbf{Avg.} \\
\midrule
AR & 0.48 $(+0.36)$ & 0.44 $(+0.13)$ & 0.46 $(+0.18)$ & 0.40 $(+0.22)$ & \cellcolor{aliceblue}0.45 $(+0.22)$ \\
DE & 0.38 $(+0.12)$ & 0.44 $(+0.17)$ & 0.42 $(+0.00)$ & 0.48 $(+0.26)$ & \cellcolor{aliceblue}0.43 $(+0.14)$ \\
EN & 0.22 $(-0.04)$ & 0.47 $(+0.12)$ & 0.22 $(-0.16)$ & 0.28 $(-0.06)$ & \cellcolor{aliceblue}0.30 $(-0.04)$ \\
FR & 0.28 $(+0.04)$ & 0.51 $(+0.28)$ & 0.40 $(-0.22)$ & 0.48 $(+0.22)$ & \cellcolor{aliceblue}0.42 $(+0.08)$ \\
IT & 0.32 $(+0.22)$ & 0.46 $(+0.23)$ & 0.36 $(+0.08)$ & 0.34 $(+0.04)$ & \cellcolor{aliceblue}0.37 $(+0.14)$ \\
JA & 0.16 $(-0.04)$ & 0.40 $(+0.05)$ & 0.32 $(-0.20)$ & 0.30 $(-0.10)$ & \cellcolor{aliceblue}0.30 $(-0.07)$ \\
PT & 0.46 $(+0.12)$ & 0.41 $(+0.11)$ & 0.46 $(-0.26)$ & 0.32 $(+0.04)$ & \cellcolor{aliceblue}0.41 $(+0.00)$ \\
ZH & 0.22 $(+0.08)$ & 0.32 $(+0.13)$ & 0.38 $(-0.04)$ & 0.22 $(-0.20)$ & \cellcolor{aliceblue}0.29 $(-0.01)$ \\
\midrule
\rowcolor{aliceblue}
\textbf{Avg.} 
& 0.32 $(+0.11)$ 
& 0.43 $(+0.15)$ 
& 0.38 $(-0.08)$ 
& 0.35 $(+0.05)$ 
& 0.37 $(+0.06)$ \\
\bottomrule
\end{tabular}
}
\label{tab:qwen_fixed_patch_31_3}
\end{table*}

\subsection{Optimal Configurations and Iterative Results}
\label{appdx:optimal}
To identify effective intervention pathways, we search over window sizes $w \in \{1, 3, 5\}$. The intervention uses direct activation replacement, with no additive mixing or additional scaling. We evaluate only valid source-destination pairs whose layer windows remain within model bounds. Overlapping windows are allowed as long as the source layer is deeper than the destination layer. \Cref{tab:backpatching_best_config} catalogs the best valid configuration for each model, language, and task setting. While the exact hyperparameter values exhibit slight variations depending on the linguistic and task context, the selected interventions often involve routing representations from late layers back to early layers. Note that to maintain computational efficiency, these configurations were not derived using the entire dataset, but were instead identified by evaluating a randomly selected subset of 50 samples per task and language. Final performance reported in the main back-patching results is evaluated on a separate 100-sample set per task and language, rather than on this search subset.

To further analyze intervention pathways, we performed iterative back-patching across all hidden layers. \cref{fig:iterative_patching} illustrates that applying the back-patching intervention iteratively does not strictly correlate with continuous improvements in audio accuracy. In many cases, a single back-patching iteration yields the largest performance gain, with subsequent iterations offering diminishing returns or slight regressions.

\subsection{Evaluation on Natural Voice}
\label{appdx:natural_voice_backpatching}
While the primary causal interventions were conducted using length-calibrated TTS data to ensure symmetric evaluations across both conflict directions (i.e., $A_{\text{base}} \rightarrow T_{\text{CF}}$ and $A_{\text{CF}} \rightarrow T_{\text{base}}$), it is also important to examine whether the identified mechanism extends to realistic acoustic variations. \cref{tab:backpatching_natual_voice} presents the back-patching performance using natural human speech datasets. Due to the inherent design of the original benchmark, this evaluation is restricted to the $A_{\text{CF}} \rightarrow T_{\text{base}}$ conflict direction, since matched natural-voice samples are only available for that direction. We therefore do not claim symmetric natural-voice generalization across both conflict directions. 

The results indicate that back-patching can improve audio reliance under the acoustic variability of natural voice, but the effect is model-dependent. The gains are stronger for Qwen, while Ultravox shows weaker average improvements. Overall, these results suggest that the intervention can reduce text dominance beyond synthetic TTS settings, while leaving broader naturalistic generalization as an open direction.

\subsection{Behavior under Aligned (Non-Conflict) Scenarios}
\label{appdx:aligned_backpatching}
A robust causal intervention should selectively resolve modality conflicts without degrading performance under aligned inputs. To validate this, we applied back-patching to aligned data, where both the audio and text modalities present identical, correct information.

As shown in \cref{tab:backpatching_aligned}, the intervention does not substantially degrade performance under aligned inputs. The performance deltas ($\Delta$) across all models and tasks hover around zero ($\approx 0.00$). This suggests that targeted back-patching mainly affects modality competition under conflict, without noticeably disrupting standard processing when the modalities are already in agreement.

\begin{table*}[!t]
\centering
\caption{\textbf{Attention mass distribution and directional shifts during the final generation step.} The table reports the baseline attention mass directed toward text and audio tokens (normalized by token length) and the absolute shift ($\Delta$) following the back-patching intervention. The positive $\Delta$ Audio values suggest increased attention toward audio tokens after intervention.}
\label{tab:dynamics_attn_backpatching}
\setlength{\tabcolsep}{10pt}
\resizebox{\textwidth}{!}{%
\begin{tabular}{ll c c c c c}
\toprule
\textbf{Model} & \textbf{Task} & 
\makecell{\textbf{Attn. to Text}} & 
\makecell{\textbf{Attn. to Audio}} & 
\makecell{\textbf{Mean }$\bm{\Delta}$ \textbf{Text}} & 
\makecell{\textbf{Mean }$\bm{\Delta}$ \textbf{Audio}} & 
\makecell{\textbf{Mean }$\bm{\Delta}$ \textbf{Other}} \\
\midrule
\multirow{5}{*}{Qwen} 
& Adjective Swap  & 0.0424 & 0.3260 & $-0.0052$ & $+0.1419$ & $+0.0049$ \\
& Negation Swap  & 0.0372 & 0.2102 & $-0.0079$ & $+0.0310$ & $-0.0020$ \\
& Number Swap  & 0.0465 & 0.2623 & $-0.0021$ & $+0.0805$ & $+0.0026$ \\
& Time Swap  & 0.0415 & 0.3172 & $-0.0046$ & $+0.1332$ & $+0.0063$ \\
\cmidrule(lr){2-7}
& Avg. & 0.0419 & 0.2789 & $-0.0050$ & $\mathbf{+0.0967}$ & $+0.0030$ \\
\midrule
\multirow{5}{*}{Ultravox} 
& Adjective Swap  & 0.0022 & 0.0065 & $+0.0002$ & $-0.0004$ & $-0.0004$ \\
& Negation Swap  & 0.0018 & 0.0058 & $-0.0000$ & $-0.0002$ & $-0.0000$ \\
& Number Swap  & 0.0060 & 0.0167 & $+0.0042$ & $+0.0097$ & $-0.0019$ \\
& Time Swap  & 0.2435 & 0.4849 & $+0.2416$ & $+0.4779$ & $+0.0046$ \\
\cmidrule(lr){2-7}
& Avg. & 0.0634 & 0.1285 & $+0.0615$ & $\mathbf{+0.1218}$ & $+0.0006$ \\
\bottomrule
\end{tabular}%
}
\vspace{10pt}
\end{table*}

\FloatBarrier

\subsection{Activation Dynamics of Modality Restoration}
\label{appdx:dynamics_backpatching}

To understand how back-patching affects text dominance, we analyze the internal activation dynamics. Specifically, we measure shifts in hidden state magnitudes (L2 norm) and attention weights to examine whether the acoustic signal is amplified and prioritized.

\noindent\textbf{Signal Amplification via L2 Norm.} 
In the residual stream, the L2 norm serves as a proxy for representation strength. \cref{tab:dynamics_L2_backpatching} shows that back-patching increases the L2 norm of audio tokens across tasks. This increase is larger in Qwen (avg. $\Delta \text{L2} = +81.23$) and more moderate in Ultravox ($+10.59$), aligning with their different accuracy gains. This suggests that injecting late-layer representations amplifies acoustic features, which may help counteract their suppression by the textual pathway.

\begin{table}[!t]
\centering
\small
\caption{\textbf{L2 norm measurements and representation shift deltas for audio tokens across tasks.} The table reports the magnitude of audio representations after back-patching (L2 After) and the change compared to the baseline ($\Delta$ L2). The positive $\Delta$ L2 values indicate increased acoustic representation magnitude after the intervention.}\label{tab:dynamics_L2_backpatching}
\setlength{\tabcolsep}{8pt}
\resizebox{\linewidth}{!}{%
\begin{tabular}{ll c c}
\toprule
\textbf{Model} & \textbf{Task} & \makecell{\textbf{L2 (After)}} & \makecell{$\bm{\Delta}$ \textbf{L2}} \\
\midrule
\multirow{5}{*}{Qwen} 
& Adjective Swap  & 243.64 & $+65.99$ \\
& Negation Swap   & 337.27 & $+159.85$ \\
& Number Swap     & 217.27 & $+39.80$ \\
& Time Swap       & 236.45 & $+59.26$ \\
\cmidrule(lr){2-4}
& Avg.   & 258.66 & $\mathbf{+81.23}$ \\
\midrule
\multirow{5}{*}{Ultravox} 
& Adjective Swap  & 47.67  & $+3.42$ \\
& Negation Swap   & 44.24  & $-0.06$ \\
& Number Swap     & 49.99  & $+5.83$ \\
& Time Swap       & 77.35  & $+33.17$ \\
\cmidrule(lr){2-4}
& Avg. & 54.81  & $\mathbf{+10.59}$ \\
\bottomrule
\end{tabular}%
}
\end{table}

\noindent\textbf{Routing Reallocation via Attention Shifts.} 
For the amplified signal to impact the output, the model must route it during generation. \cref{tab:dynamics_attn_backpatching} shows that back-patching increases attention toward audio tokens ($\Delta \text{Audio} > 0$) on average across both models. In Qwen, this audio increase (avg. $+0.0967$) is accompanied by reduced text attention (avg. $-0.0050$), suggesting a partial reallocation from text to audio. 

Together, these metrics suggest a two-step mechanism: the intervention first \textit{amplifies} the acoustic signal (L2 norm), which is then accompanied by greater attention to audio information during the final prediction.

\subsection{Intervention Heatmaps Across Diverse Languages}
\label{appdx:heatmaps}
\begin{figure*}[t]
    \centering
    
    \begin{subfigure}[b]{0.48\textwidth}
        \centering
        \includegraphics[width=\textwidth]{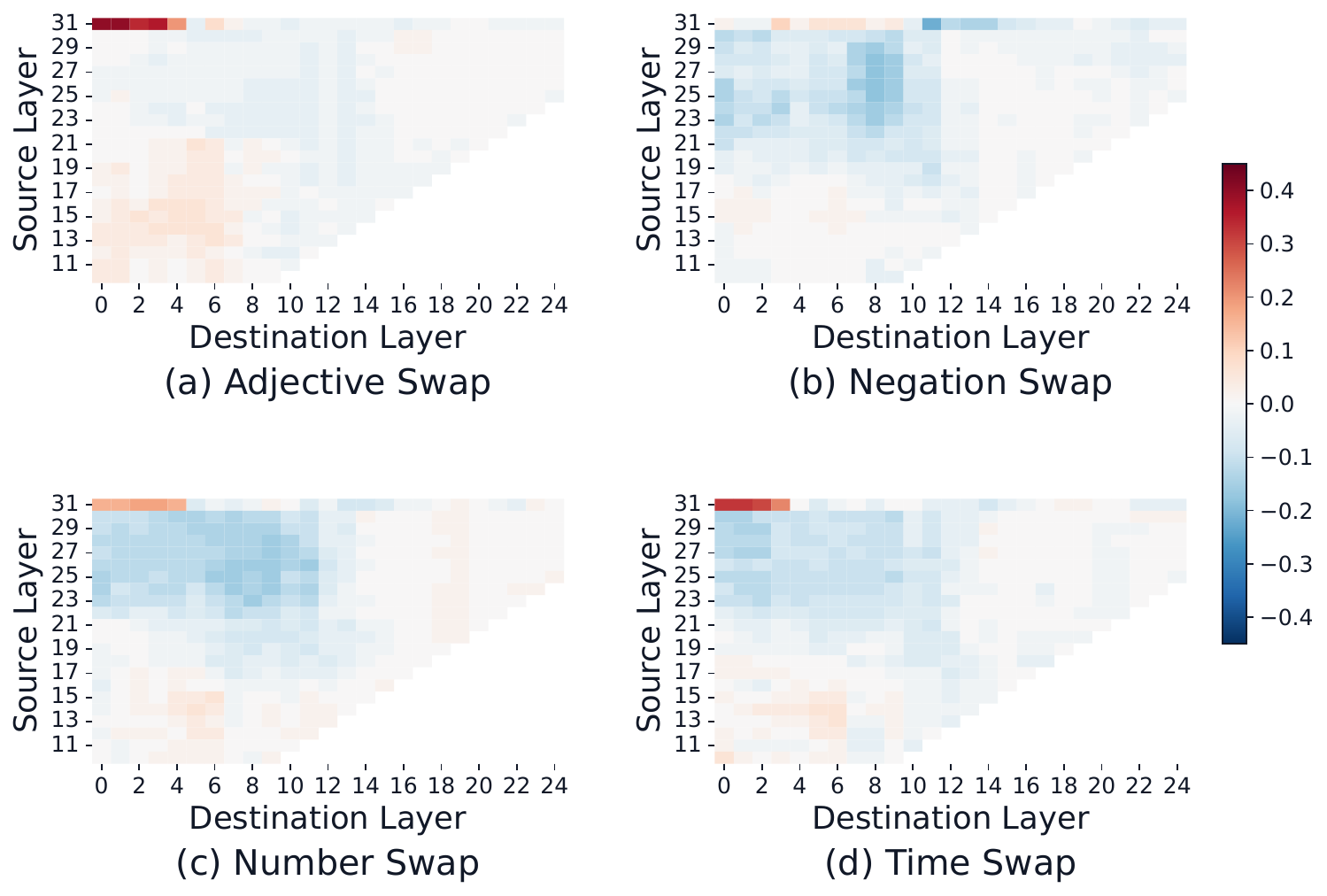} 
        \caption{Arabic (AR)}
        \label{fig:heatmap_AR}
    \end{subfigure}
    \hfill 
    \begin{subfigure}[b]{0.48\textwidth}
        \centering
        \includegraphics[width=\textwidth]{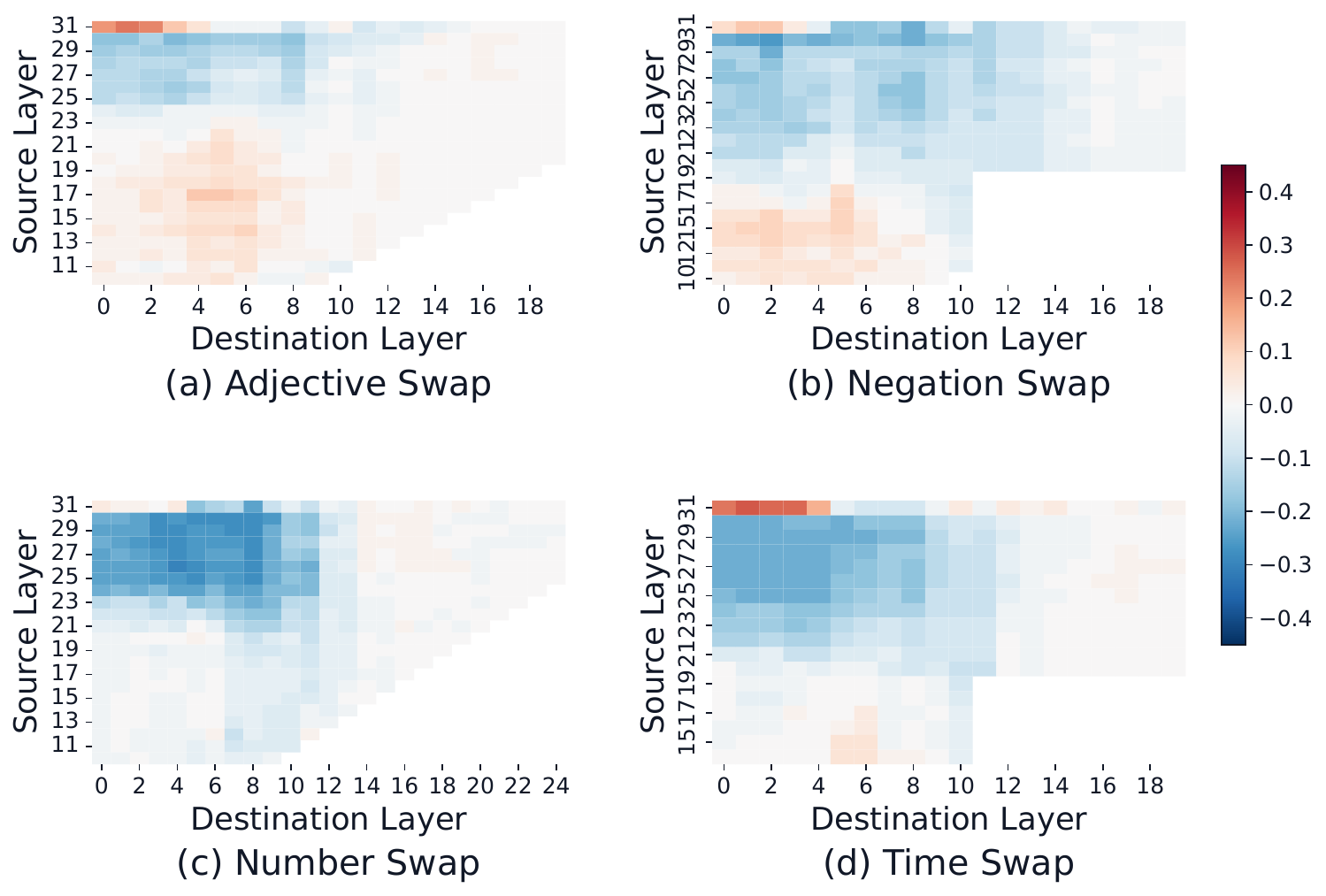} 
        \caption{German (DE)}
        \label{fig:heatmap_DE}
    \end{subfigure}
    
    \vspace{1em} 
    
    \begin{subfigure}[b]{0.48\textwidth}
        \centering
        \includegraphics[width=\textwidth]{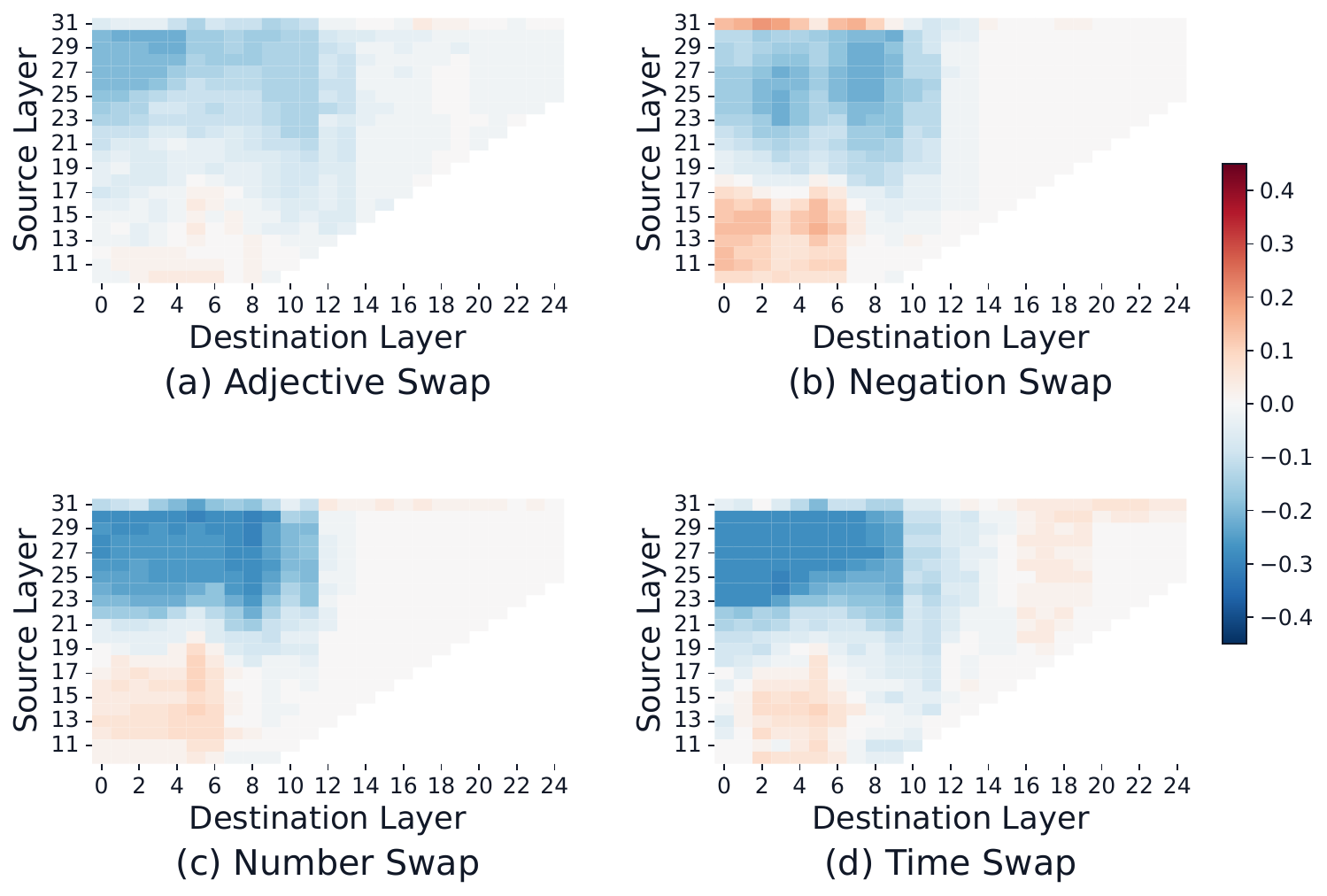} 
        \caption{English (EN)}
        \label{fig:}
    \end{subfigure}
    \hfill
    \begin{subfigure}[b]{0.48\textwidth}
        \centering
        \includegraphics[width=\textwidth]{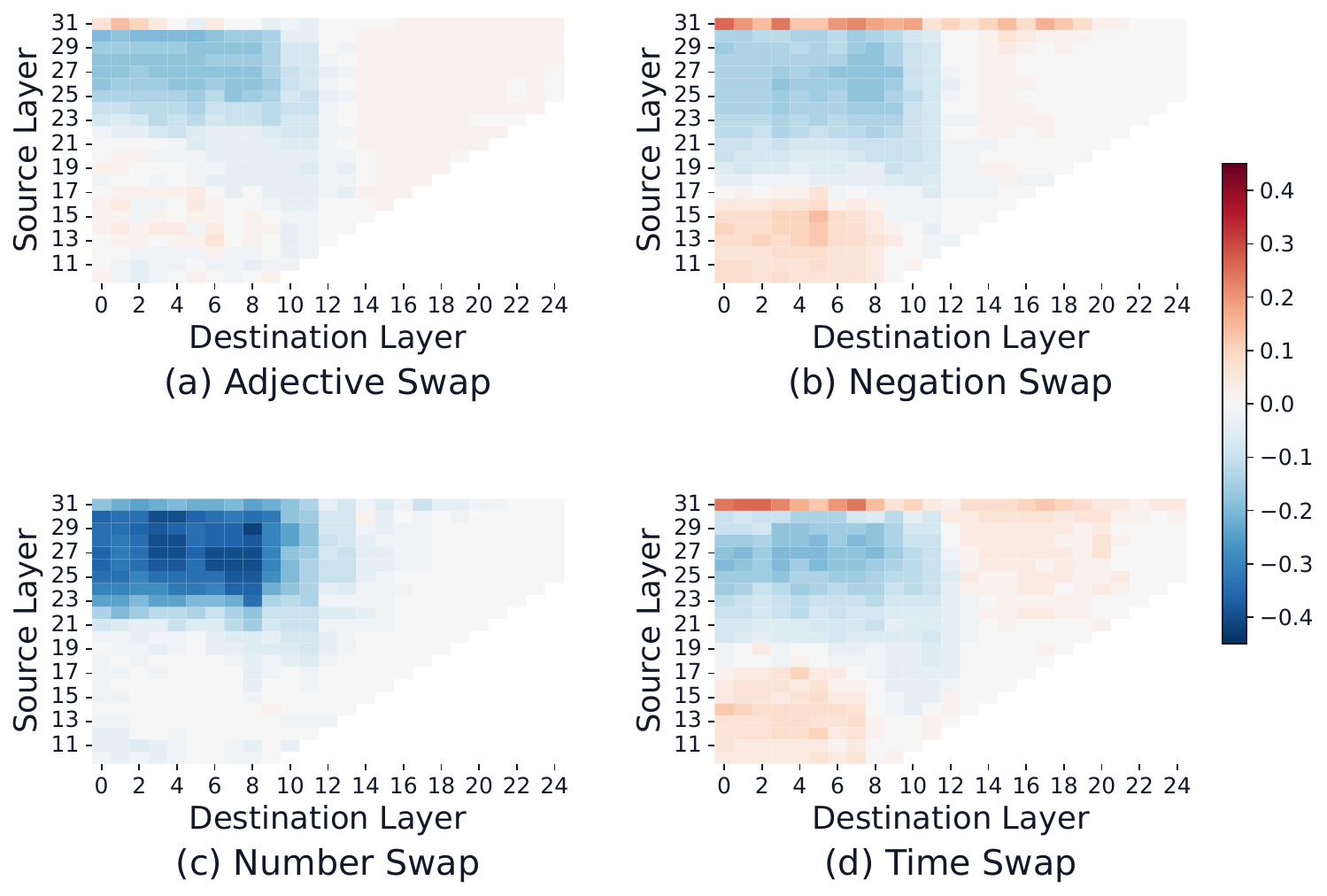} 
        \caption{French (FR)}
        \label{fig:heatmap_FR}
    \end{subfigure}
    
    \caption{\textbf{Intervention heatmaps across a linguistically diverse subset of languages.} The plots illustrate the back-patching results for Arabic (AR), German (DE), English (EN), and French (FR). Despite minor topological variations, the main trend shows stronger performance when routing representations from the late layers back to the early layers.}
    \label{fig:backpatching_heatmaps_diverse}
\end{figure*}

To examine whether the spatial distribution of multimodal circuits is similar across languages, \cref{fig:backpatching_heatmaps_diverse} presents intervention heatmaps for Arabic (AR), German (DE), English (EN) and French (FR). For computational efficiency, these visualizations were computed on a random 50-sample subset per task. 

Despite minor topological variations across languages and tasks, the main trend is similar: performance gains tend to localize in regions where representations are patched from the final layers back to the early layers. This suggests that late-to-early modality routing is a recurring pattern in our evaluated settings rather than an artifact of a single language.

\begin{table}[!t]
\centering
\small
\caption{\textbf{Licenses of datasets, models, and codebases used in this work.} The table details the open-source licenses and official repository links for the ALME benchmark, the evaluated Audio LLMs, and the referenced methodology.}
\resizebox{\columnwidth}{!}{
\begin{tabular}{llcl}
\toprule
\textbf{Type} & \textbf{Asset} & \textbf{License} & \textbf{Source} \\
\midrule
Dataset & ALME Benchmark & Apache 2.0 & \href{https://github.com/jb1999/alme-benchmark/tree/main}{GitHub} \\
\midrule
Model & Qwen2-audio & MIT & \href{https://huggingface.co/Qwen/Qwen2-Audio-7B-Instruct}{Huggingface} \\
      & Ultravox & MIT & \href{https://huggingface.co/fixie-ai/ultravox-v0_6-llama-3_1-8b}{Huggingface} \\
\midrule
Codebase & VLM Circuits Analysis & Unspecified & \href{https://github.com/technion-cs-nlp/vlm-circuits-analysis}{GitHub} \\
\bottomrule
\end{tabular}
}
\label{tab:licenses}
\end{table}

\section{AI Assistant Usage Statement}

During the preparation of this manuscript, we utilized ChatGPT and Gemini strictly for linguistic refinement, including grammatical corrections and stylistic improvements. All scientific hypotheses, experimental analyses, and core intellectual contributions remain entirely the original work of the human authors.


\section{Licenses of Datasets and Models}
\label{sec:asset_licenses}

We summarize the licenses of all datasets and pretrained models used in this work in Table~\ref{tab:licenses}. 
All assets are used in accordance with their respective licenses.

    
    
    
    

\end{document}